\documentclass[paper]{aa} 
\usepackage{multirow} 
\usepackage{natbib}
\usepackage{amssymb,amsmath}
\usepackage{graphicx, subfigure}
\usepackage{siunitx}
\usepackage{txfonts}
\usepackage{url}
\usepackage{threeparttable}
\usepackage[table]{xcolor}
\usepackage{makecell}
\usepackage{epstopdf}
\usepackage{color, colortbl}

\definecolor{Gray}{gray}{0.9}

\begin{document}

   \title{Inner dusty envelope of the AGB stars W~Hydrae, SW\,Virginis, and R~Crateris using SPHERE/ZIMPOL}

   \author{T. Khouri\inst{1}\thanks{{\it Send offprint requests to T. Khouri}\newline \email{theo.khouri@chalmers.se}},
W. H. T. Vlemmings\inst{1}, C. Paladini\inst{2}, C. Ginski \inst{3,4}, E. Lagadec \inst{5}, M. Maercker\inst{1},
P. Kervella \inst{6}, E. De Beck \inst{1}, L. Decin \inst{7}, A. de Koter \inst{3,7}, L. B. F. M. Waters \inst{3,8}
}

\institute{Department of Space, Earth and Environment, Chalmers University of Technology, Onsala Space Observatory, 439 92 Onsala, Sweden 
\and European Southern Observatory, Alonso de Cordova 3107, Vitacura, Santiago, Chile
\and Astronomical Institute ``Anton Pannekoek", University of Amsterdam, PO Box 94249, 1090 GE Amsterdam, The Netherlands 
\and    Sterrewacht Leiden, P.O. Box 9513, Niels Bohrweg 2, 2300 RA Leiden, The Netherlands 
\and Laboratoire Lagrange, Universit\'e C\^ote d'Azur, Observatoire de la C\^ote d'Azur, CNRS, Blvd de l'Observatoire, CS 34229, 06304 Nice cedex 4, France 
\and LESIA, Observatoire de Paris, Universit\'e PSL, CNRS, Sorbonne Universit\'e, Univ. Paris Diderot, Sorbonne Paris Cit\'e, 5 place Jules Janssen, 92195 Meudon, France 
\and Instituut voor Sterrenkunde, KU Leuven, Celestijnenlaan 200D B-2401, 3001 Leuven, Belgium 
\and SRON Netherlands Institute for Space Research, Sorbonnelaan 2, 3584 CA Utrecht, The Netherlands 
}

  \abstract
  {The asymptotic giant branch (AGB) marks the final evolutionary stage of stars with initial masses between $\sim0.8$ and 8~$M_\odot$. During this phase, stars undergo copious mass loss. The well-accepted mass-loss mechanism requires radiation pressure acting on dust grains that form in the density-enhanced and extended AGB stellar atmospheres. The details of the mass-loss process are not yet well understood, however.}
 {We aim to study the spatial distribution and properties of the first grains that form around AGB stars.}
 {Using the extreme-adaptive-optics imager and polarimeter SPHERE/ZIMPOL, we observed light polarised by grains around the AGB stars W\,Hya, SW\,Vir, and R\,Crt, with mass-loss rates between 10$^{-7}$ and 10$^{-6}~M_\odot~{\rm yr^{-1}}$.}
 {We find the distribution of dust to be asymmetric around the three targets. A biconical morphology is seen for R Crt, with a position angle that is very similar to those inferred from interferometric observations of maser emission and of mid-infrared continuum emission. The cause of the biconical outflow cannot be directly inferred from the ZIMPOL data. The dust grains polarise light more efficiently at 0.65~$\mu$m for R\,Crt and SW\,Vir and at 0.82~$\mu$m for W\,Hya. This indicates that at the time of the observations, the grains around SW\,Vir and R\,Crt had sizes $< 0.1~\mu$m, while those around W\,Hya were larger, with sizes $\gtrsim 0.1~\mu$m. The asymmetric distribution of dust around R\,Crt makes the interpretation more uncertain for this star, however. We find that polarised light is produced already from within the visible photosphere of W~Hya, which we reproduce using models with an inner dust shell that is optically thick to scattering. The radial profile of the polarised light observed around W\,Hya reveal a steep dust density profile. We find the wind-acceleration region of W\,Hya to extend to at least $\sim 7~R_\star$, in agreement with theoretical predictions of wind acceleration up to $\sim 12~R_\star$.}
{}
               
\titlerunning{Semi-regular variable stars W\,Hya, SW\,Vir, and R\,Crt studied with SPHERE/ZIMPOL}
\authorrunning{T. Khouri et al.}

\maketitle
%

\section{Introduction}

At the end of their lives, stars with  masses between 0.8 and 8~$M_\odot$ present strong mass loss during their evolution on the asymptotic giant branch (AGB).  This plays an important role in the enrichment of the interstellar medium \citep{Habing2003}. The mass-loss
mechanism is widely accepted to be a two-step process in which stellar pulsations and/or convective motion increase the
density scale height of the atmosphere, and radiation pressure acting on newly formed dust grains drives the wind \citep{Hoefner2018}.
The details of this process are not yet well understood, however, and predicting the mass-loss rate of an AGB star from first principles
is not possible at present. For instance, the relative contribution between convection and stellar pulsations in creating the extended AGB atmospheres is not known,
and neither is how this balance is affected by stellar properties and AGB evolutionary phase. Moreover, the dust-formation process is complex,
and a detailed modelling of all relevant chemical reactions in the highly dynamical AGB atmospheres is still in its infancy. For oxygen-rich
AGB stars (in which the atmospheric carbon-to-oxygen ratio is lower than one), the scenario is particularly complex because opaque dust
species that could drive the wind through absorption of photons do not form close enough to the star to trigger the outflow \citep{Woitke2006}.
Therefore, the outflows of O-rich AGB stars have been proposed to be driven by radiation pressure caused by scattering of visible and near-infrared light on
relatively large ($\gtrsim 0.1~\mu$m) but translucent dust grains \citep{Hofner2008}.
High-angular-resolution observations of polarised light toward O-rich AGB stars have revealed that grains with sizes $\gtrsim 0.1~\mu$m indeed
exist within a few stellar radii from the central star \citep{Norris2012,Kervella2015,Khouri2016,Ohnaka2016,Ohnaka2017}.

The nature of the first dust grains to form around O-rich AGB stars is still debated both from an observational
\citep[e.g. ][]{Zhao-Geisler2012,Karovicova2013,Khouri2015,Kaminski2016,Kaminski2017}
and a theoretical perspective \citep[e.g. ][]{Gail2016,Gobrecht2016,Hoefner2016,Bladh2017}.
Nonetheless, observations indicate that aluminium-oxide grains form
closer to the star at approximately two stellar radii, while opaque silicates form farther away, at a few times this distance \citep[e.g. ][]{Zhao-Geisler2012,Karovicova2013}.
Theoretical models indicate that aluminium oxide might indeed form directly from the gas phase \citep{Gobrecht2016}, but the sizes and
the lattice structure of such grains are still a matter of discussion \citep[e.g., ][]{Decin2017}.
Whether aluminium oxide \citep{Gobrecht2016}, titanium oxide \citep{Plane2013}, or silicates \citep{Gail2016} are the first species to condense from the gas phase
is also still debated. Constraints on the composition of the dust have been actively pursued recently
\citep[e.g. ][]{Kaminski2016,Kaminski2017,Decin2017,Khouri2018}, but no conclusion has been reached.
Nonetheless, the condensation of some type of silicate grain is deemed necessary for the development of the outflows of O-rich AGB stars because of the relatively
low abundance of aluminium \citep[e.g. ][]{Bladh2012}.
A possible scenario is that the first aluminium-oxide grains
act as seeds for the subsequent
condensation of iron-free silicates \citep[e.g. ][]{Kozasa1997,Hoefner2016}, and that iron is incorporated farther out in the outflow \citep[e.g. ][]{Bladh2015}.
The way in which all these dust properties vary for AGB stars with different characteristics (such as pulsation amplitude and mode, period, and mass-loss rate)
is not known either.

In this context, we observed polarised light towards three semi-regular-variable O-rich AGB stars
using the Zurich Imaging Polarimeter (ZIMPOL), an extreme adaptive-optics imager and polarimeter of the Spectro-Polarimetric High-contrast Exoplanet Research (SPHERE) on the Very Large Telescope (VLT).
In this way, we aim to study the grains expected to drive the outflows using the polarised light they produce through scattering.
From these observations, we can study their spatial distribution, and in particular, the radial distance at which they from.

\section{Observations}

\subsection{Description of the observed stars}

The three stars we selected are W\,Hya, SW\,Vir, and R\,Crt. These are
semi-regular variable O-rich AGB stars that present mass-loss rates ranging from $\sim 10^{-7}$~$M_\odot{\rm yr^{-1}}$ to
$\sim 10^{-6}$~$M_\odot{\rm yr^{-1}}$. 
We note that ZIMPOL observations of AGB stars indicate strong variability on timescales of one month \citep{Khouri2016,Ohnaka2017}, and the
observed differences between sources based on single-epoch observations should therefore be interpreted with care and need to be followed up.

\subsubsection{W\,Hya}

At a distance of $\approx 100$~pc \citep{Vlemmings2003,vanLeeuwen2007}, W Hya is the second brightest star in the K band and one of the brightest infrared sources in the sky \citep{Wing1971}.
Its visual magnitude varies between $\sim 6$~mag and 9~mag in about 388 days \citep{Zhao-Geisler2011}.
A mass-loss rate between $10^{-7}$ and $1.5 \times 10^{-7}~M_\odot$~yr$^{-1}$ is found based on models for CO emission lines \citep[e.g.][]{Maercker2008,Khouri2014}, while a main-sequence mass between 1.0 and 1.5~$M_\odot$
has been derived from studies of oxygen isotopic ratios \citep{Khouri2014a,Danilovich2017}.
The stellar radius varies in time and strongly with wavelength \citep{Zhao-Geisler2011}.
To our knowledge, the smallest radius of $\sim 15$~mas was reported by \cite{Woodruff2009} based on observations
in the near-infrared at pulsation phase 0.8. Throughout the paper, we consider a reference stellar radius ($R_\star)$
of 18~mas as measured by \citeauthor{Woodruff2009} at pulsation phase 0.54, which is similar to when the observations presented by us were acquired (0.5).

Observations of scattered light using the Nasmyth Adaptive Optics System Near-Infrared Imager and Spectrograph (NaCo) on the VLT in March 2009 and June 2010 \citep{Norris2012} and
using ZIMPOL in July 2015 \citep{Ohnaka2016} and March 2016 \citep{Ohnaka2017}
show a dust envelope containing large dust grains (with radii $\gtrsim 0.1~\mu$m) with an inner radius of 25 mas \citep{Ohnaka2016}, 34 mas \citep{Ohnaka2017} and 37 mas \citep{Norris2012}
at phases 0.92, 0.54, and 0.2, respectively. 
Whether the grains observed through scattered light are aluminium oxide or silicates is still unclear \citep[e.g. ][]{Ohnaka2017}.
Interferometric infrared observations reveal that amorphous aluminium oxide emission is produced close to the star, within $\sim 50$~mas ($5$~au) in radius, 
while silicate emission arises much farther out, at $r \gtrsim 400$~mas \citep[$40$~au,][]{Zhao-Geisler2011}.
By studying the infrared spectrum of W\,Hya, \cite{Khouri2015} found that
emission from amorphous aluminium oxide must originate almost exclusively by material that is gravitationally bound to the star.

Recent images of the sub-millimeter continuum emission of W\,Hya acquired using the Atacama Large Millimeter/submillimeter Array (ALMA) at pulsation phase 0.3 (March 2015) revealed an asymmetric source with size $50.9 \times 56.5$~mas and an unresolved hotspot on the surface.
The same data show that the extended atmosphere of W\,Hya, traced by the $v=1, J=3-2$ CO line, extends
out to $\approx 48$~mas ($4.8$~au) in radius \citep{Vlemmings2017}, tracing the extended atmosphere.
The grains that produce the scattered light might therefore also be present in the high-gas-density environment of the extended atmosphere that is gravitationally bound to the star.
An independent analysis of these ALMA observations suggests aluminium to be highly depleted into dust in the inner regions and silicon not to be significantly depleted \citep{Takigawa2017}.
We note that the simplified model employed by \citeauthor{Takigawa2017} is most likely not applicable to the innermost regions of the circumstellar envelope probed by the ALMA observations,
and their results must be confirmed by more detailed calculations.

\subsubsection{R\,Crt}

R\,Crt is a semi-regular variable (type SRb) with a pulsation period of 160 days and
a visual magnitude amplitude of 1.4~mag \citep{Jura1992}. Its reported distance varies from $\sim 170$~pc \citep{Bowers1994} to $261^{+86}_{-42}$~pc \citep{vanLeeuwen2007}.
Based on observations of low-excitation CO lines, \cite{Olofsson2002} found R\,Crt to have
the highest mass-loss rate ($\sim 8 \times 10^{-7}$~M$_\odot$/yr) of the 43 nearby semi-regular variables in their sample,
with an expansion outflow velocity of 10.6~km/s.
A diameter -- V-K relation \citep{vanBelle1999} predicts a K-band diameter of 19 mas. However,
a fit to archive VINCI (on the VLT interferometer) data using a uniform disc source model
provides a slightly smaller solution, with a K-band diameter of about $14.8\pm0.2$ mas.
Observations of R\,Crt using MIDI on the VLTI in the mid-infrared and at short baselines (with a resolution of $\sim 68$~mas)
are best fit using an elliptical Gaussian model with an axis ratio of 0.7 and a position angle of 157$^\circ$ \citep{Paladini2017}. At the wavelengths
of the MIDI observations, circumstellar dust becomes a significant source of opacity. This means that the dust distribution might affect the measurement,
but the extent to which this occurs is not yet clear.
In the far-infrared and on
a larger scale, a dust shell with radius of a few arcminutes is seen both in IRAS \citep{Young1993} and HERSCHEL/PACS \citep{Cox2012} images.
This shell is probably created by the interaction between the stellar outflow and the interstellar medium (ISM).

R\,Crt presents SiO, H$_2$O, and OH maser emission. VLBI maps of the 22~GHz water maser \citep{Ishitsuka2001} reveal
a biconical outflow at a position angle of 136$^\circ$ (measured from north to east)
at $r \lesssim 100$~mas ($\sim 5~R_\star$) from the star.
More recent simultaneous observations of SiO and H$_2$O masers show that the SiO maser spots form a ring,
which is typical for AGB stars, with a radius of $\sim 13$~mas. The H$_2$O maser
is produced almost exclusively in the south-east with respect to the star, with no emission in the north and north-west regions \citep{Kim2018}.
The OH masers were mapped by \cite{Szymczak1999}, revealing a displacement
between the red- and blue-shifted masers in the same direction as the possible biconical H$_2$O 
outflow. Additionally, the size of the OH shell inferred from these observations implies a
mass-loss rate one order of magnitude lower than that obtained from CO observations.
The magnetic field direction derived from the OH masers is in the
same direction (with position angle $\sim140^\circ$). Finally, \cite{Herpin2006} observed the polarisation
of the SiO masers and found that the magnetic field (of up to $\sim 3.7$~G) is aligned in the same position angle as the OH masers. This suggests
that the magnetic field might be related to the asymmetries seen in the star and in the dust and maser distributions.

\subsubsection{SW\,Vir}

At a distance of 143$^{+20}_{-15}$~pc \citep{vanLeeuwen2007}, SW\,Vir presents a period of $\approx 154$ days, possibly additional
periods of 164 and 1700 days \citep{Percy2001,Kiss1999}, and
a visual magnitude amplitude of $\sim 1.5$~mag \citep{Jura1992}.
The detection of Tc in the atmosphere shows that SW\,Vir is in the thermally pulsating AGB phase \citep{Lebzelter2003}.
Models fitting observations of low-excitation CO lines indicate a mass-loss rate of $4 \times 10^{-7}~M_\odot$~yr$^{-1}$
and an expansion velocity of the outflow of $7.5$~km/s \citep{Olofsson2002}.
The uniform disc diameter of SW\,Vir has been measured in the H \citep{Ridgway1982} and K \citep{Schmidtke1986} bands to be 16.8~mas and 16.7 mas, respectively. 

\begin{table}
\footnotesize
\setlength{\tabcolsep}{4pt}
\caption{Observed sources. Stellar diameters $\theta_\star$ given at visible, near-infrared, and 8.0 $\mu$m, when available.}              
\label{tab:sources}      
\centering                                      
\begin{tabular}{r c c c c c c}
\hline\hline                        
Source & $d$& $\theta_\star$ & $P$ & $V_{\rm max}$ & $\Delta V$ & $\dot{M}$\\    
& [$pc$] & [mas] & [days] & [mag] & [mag] & [$M_\odot~{\rm yr}^{-1}$]\\
W\,Hya & 98$^{\rm a, b}$ & 50$^{\rm c}$ / 41$^{\rm d}$ 
/ 80$^{\rm e}$ & 388$^{\rm e}$ & 6.0 & 4.0 & $1\times10^{-7{\rm (f)}}$ \\\
SW\,Vir & 143$^{\rm b}$ & / 16.8$^{\rm g}$ / & 150$^{\rm h}$ & 6.2 & 1.8 & $4\times10^{-7~{\rm (h)}}$ \\
R\,Crt & 260$^{\rm b}$ & / 18 / 38$^{\rm i}$& 160$^{\rm h}$ & 9.8 & 1.4 & $8\times10^{-7~{\rm (h)}}$\\
\hline                                   
\hline                                             
\end{tabular}
\tablefoot{a - \cite{Vlemmings2003}, b - \cite{vanLeeuwen2007}, c - \cite{Ireland2004}, d - \cite{Woodruff2009}, e - \cite{Zhao-Geisler2011},
f - \cite{Khouri2014}, g - \cite{White1987}, h - \cite{Olofsson2002}, i - \cite{Paladini2017}}
\end{table}

 \begin{figure*}[t]
   \centering
      \includegraphics[width= 18cm]{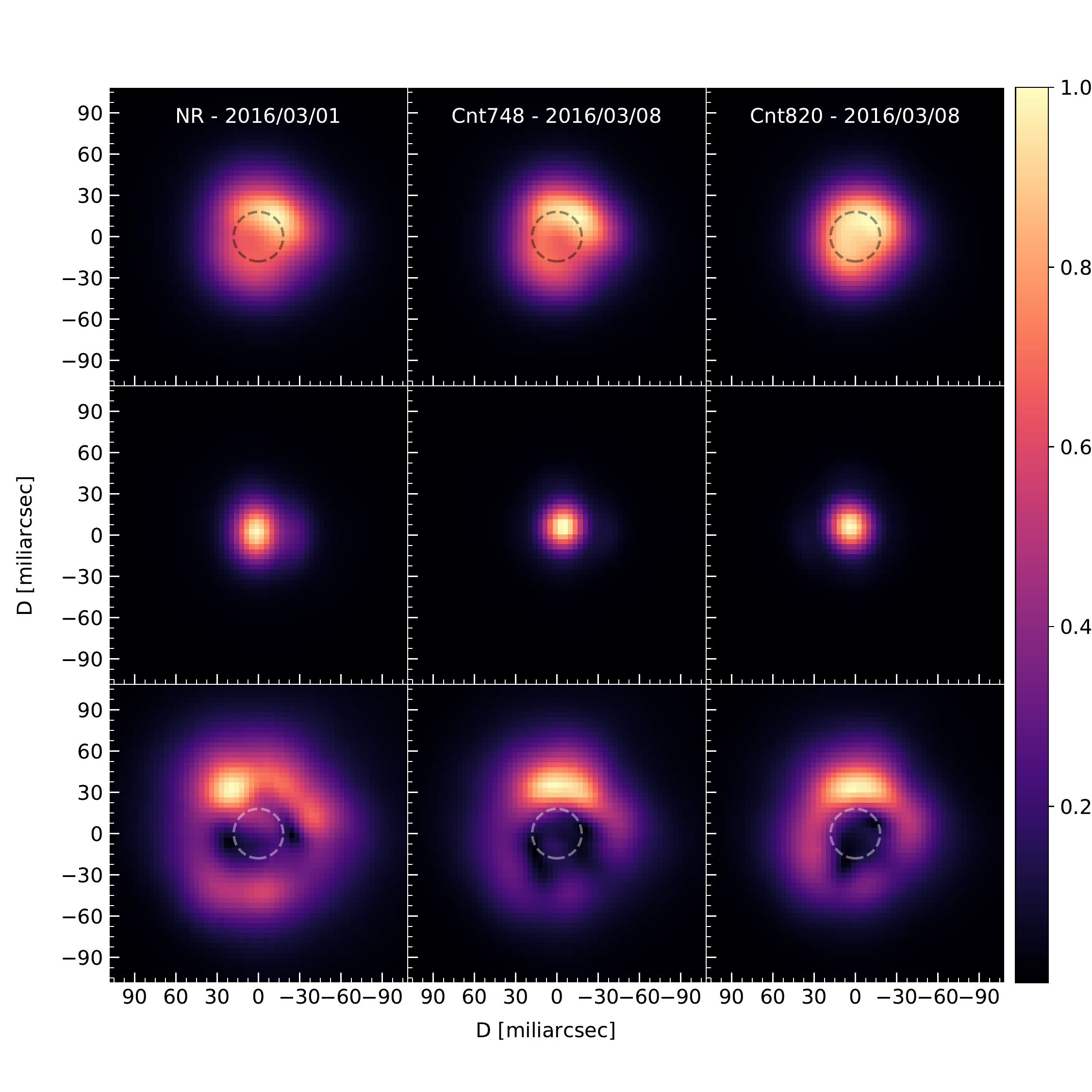}
      \caption{ZIMPOL observations of W\,Hya in filters NR, cnt748, and cnt820. {Top panels}: Total intensity of W\,Hya normalised to the peak total intensity value in the image.
      {Middle panels}: Images of the PSF-reference star  HD~123906 normalised to the peak polarised intensity.
      {Bottom panels}: Images of the polarised intensity of W\,Hya normalised to the peak polarised intensity value in the image.
      The dashed circles indicate the radius of 18~mas measured by \cite{Woodruff2009} in the near-infrared.
      }
         \label{fig:filters}
   \end{figure*}

 \begin{figure*}[t]
   \centering
      \includegraphics[width= 18cm]{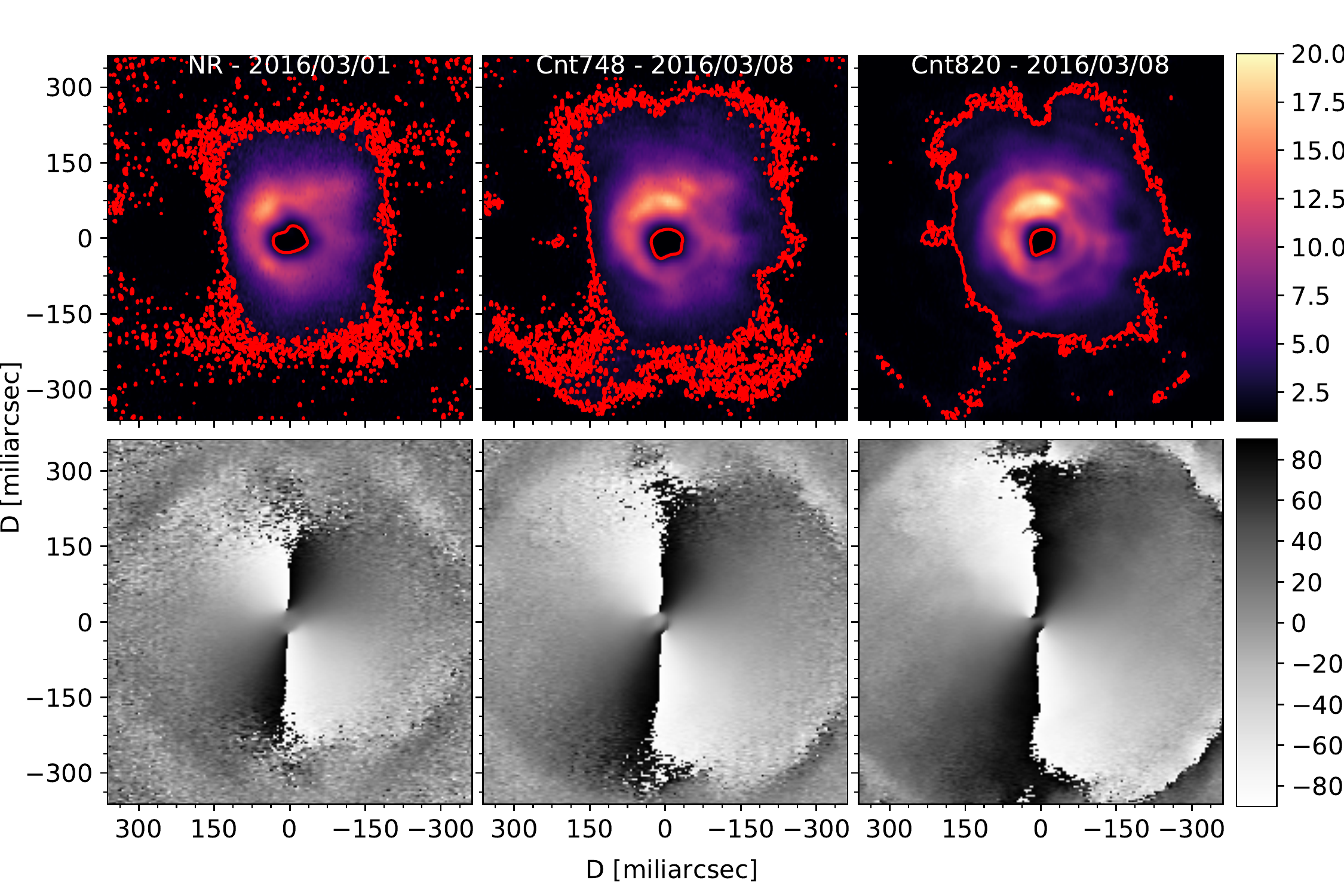}
      \caption{Polarised light in the close environment of W\,Hya in filters NR, cnt748, and cnt820.
      {Upper panels:} Colour maps showing the percentage polarisation degree observed toward W\,Hya in filters NR, cnt748, and cnt820 (from left to right).
      The red contours show the 2\% level of the polarisation degree below which instrumental polarisation can dominate.
      {Bottom panels:} Grey-scale maps showing the polarisation angle in degrees with respect to the north direction
       measured in filters NR, cnt748, and cnt820 (from left to right).
      }
         \label{fig:filtersPol}
   \end{figure*}

 \begin{figure*}[t]
   \centering
      \includegraphics[width= 18cm]{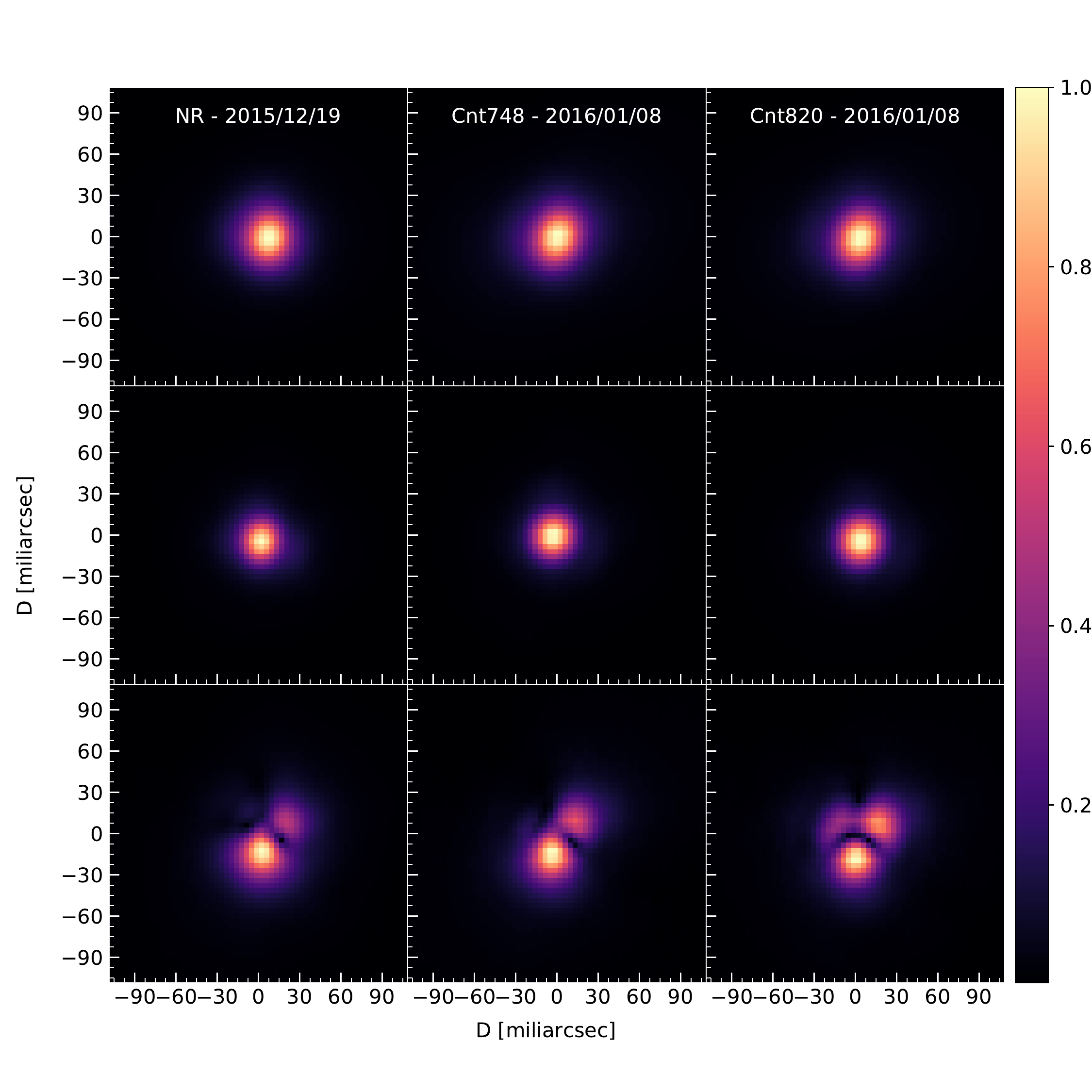}
      \caption{ZIMPOL observations of R\,Crt in filters NR, cnt748, and cnt820. 
      {Top panels}: Total intensity observed towards R\,Crt normalised to the peak intensity.
      {Middle panels}: Images of the PSF-reference star HD~96364 normalised to the peak intensity.
      {Bottom panels}: Polarised intensity observed towards R\,Crt normalised to the peak polarised intensity. The dark stripes
      seen in the centre appear in the regions where the total intensity peaks and do not correspond to physical features.
      }
         \label{fig:filtersRCrt}
   \end{figure*}

 \begin{figure*}[t]
   \centering
      \includegraphics[width= 18cm]{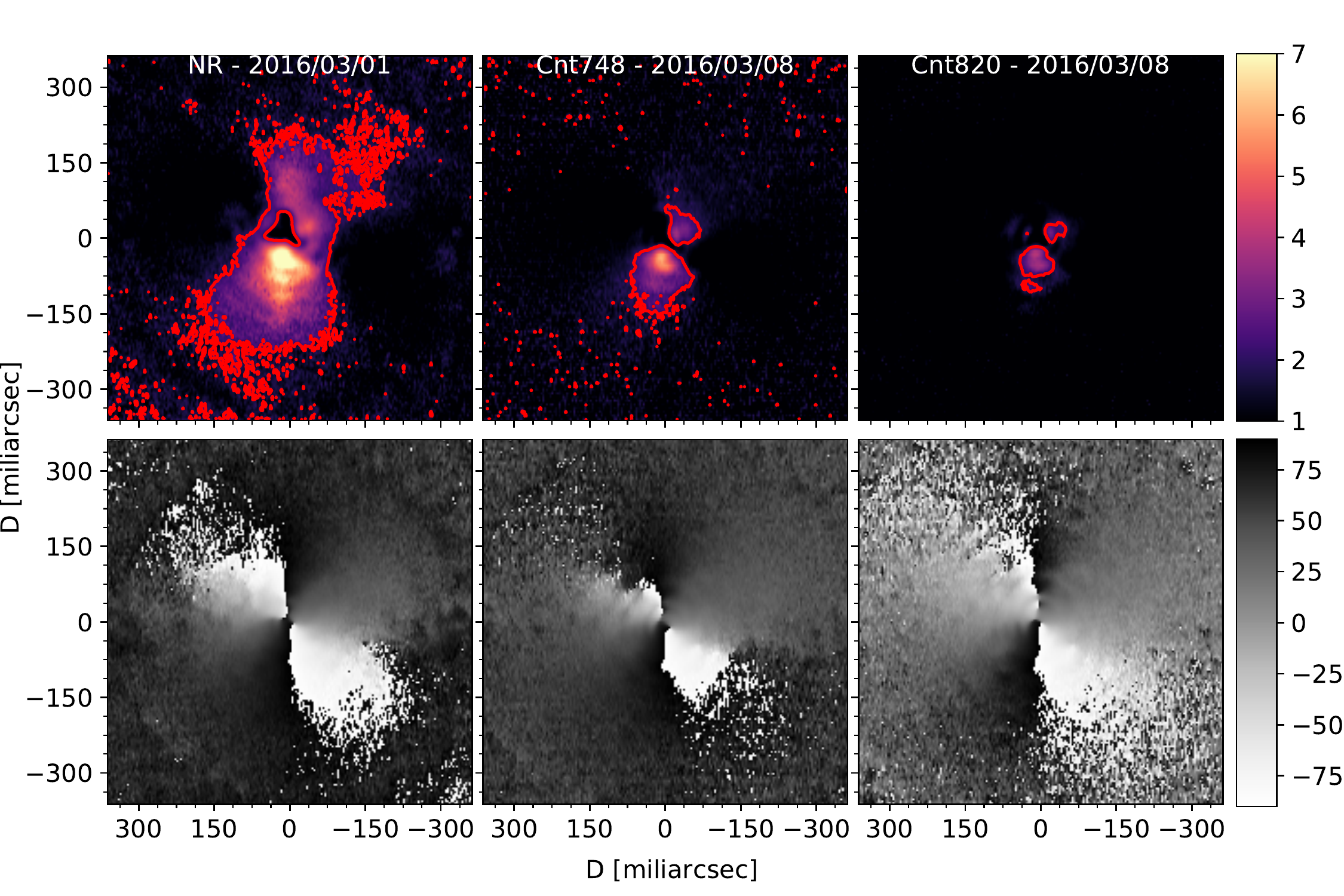}
      \caption{Polarised light in the close environment of R\,Crt in filters NR, cnt748, and cnt820.
      {Upper panels:} Colour maps showing the percentage polarisation degree observed towards R\,Crt
       in filters NR, cnt748, and cnt820 (from left to right).
      The red contours show the 2\% level of the polarisation degree below which instrumental polarisation can dominate.
      {Bottom panels:} Grey-scale maps showing the polarisation angle in degrees with respect to the north direction
       measured in filters NR, cnt748, and cnt820 (from left to right).}
         \label{fig:filtersPolRCrt}
   \end{figure*}

\begin{table*}
\footnotesize
\setlength{\tabcolsep}{4pt}
\caption{Observation log containing the filter, dates, and exposure times of observations together with other details regarding the observations and derived values.}              
\label{tab:obs}      
\centering                                      
\begin{tabular}{r r c  c@{ } c c c c c c c c@{}c@{}c}
\hline\hline                        
Source & Filter & Obs. Date & Exp time & DIT[s] & AM & $\theta$ & $S_{\rm H}$ & $S_\lambda$ & FWHM$_\star$ & FWHM$^{\rm Mod}_\star$ & Peak PD \\    
& & & [min] & x NDIT & & [$''$] & & & [$mas$] ([deg]) & [$mas$] ([deg]) & [\%] \\
\hline                                   

W\,Hya & NR & 01-Mar-16 & 32.0 & $1.2 \times 20$ & 1.09 & 0.88 -- 1.12 & 0.81 & 0.25 & $69.1 \times 71.8$ (165) & $ 63.6 \times 66.6$ (146)$^\dagger$ & 16.6 \\
  & cnt 748 & 08-Mar-16 & 24.0 & $1.2 \times 20$ & 1.03 & 0.74 -- 1.16 & 0.92 & 0.67 & $63.7 \times 69.0$ (160) & $ 58.9\times 64.4$ (159)$^\dagger$ & 18.5 \\ 
  & cnt 820 & 08-Mar-16 & 24.0 & $1.2 \times 20$ & 1.03 & 0.74 -- 1.16 & 0.92 & 0.71 & $56.4 \times 62.2$ (150) & $ 50.2 \times 56.3$ (148)$^\dagger$ & 20.2 \\ 
 \rowcolor{Gray}
 HD 123906 & NR & 01-Mar-16 & 8.6 & $1.2 \times 18$ & 1.1 & 0.98 -- 1.12 & 0.68 & 0.08 & $36.7 \times 40.1$ \phantom{17}(9) & & \\
 \rowcolor{Gray}
 HD 118877 & cnt 748 & 08-Mar-16 & 5.8 & $1.2 \times 12$ & 1.1 & 0.9 -- 1.15 & 0.72 & 0.20 & $27.3 \times 30.8$ (173) & & \\
\rowcolor{Gray}
 & cnt820 & 08-Mar-16 & 5.8 & $1.2 \times 12$ & 1.1 & 0.9 -- 1.15 & 0.72 & 0.26 & $28.0 \times 31.8$ (173) & & \\
SW\,Vir & NR & 09-Mar-16 & 34.6 & $1.2 \times 18$ & 1.1 & 0.76 -- 0.88 & 0.90 & 0.50 & $29.1 \times 31.3$ \phantom{7}(10) & $12.1 \times 16.3$ (117) & 9.5 \\ 
& cnt 748 & 10-Mar-16 & 15.4 & $1.2 \times 16$ & 1.12 & 0.77 -- 0.91 & 0.90 & 0.60 & -- & -- & 8.5 \\
& cnt 820 & 10-Mar-16 & 15.4 & $1.2 \times 16$ & 1.12 & 0.77 -- 0.91 & 0.90 & 0.65 & -- & -- & 7.0 \\
& {\it cnt 748} & 10-Mar-16 & 17.9 & $2.8 \times 16$ & 1.17 & 1.15 -- 1.93 & 0.89 & 0.57 & $28.2 \times 29.7$ (175) & $14.8 \times 16.1$ (174) & 8.1 \\
& {\it cnt 820} & 10-Mar-16 & 17.9 & $2.8 \times 16$ & 1.17 & 1.15 -- 1.93 & 0.89 & 0.62 & $28.2 \times 29.8$ (173) & $12.2 \times 13.5$ (173) & 6.4 \\ 
\rowcolor{Gray}
p\,Vir &  NR & 09-Mar-16 & 9.6 & $4.0 \times 6\phantom{0}$ & 1.1 & 1.03 -- 1.18 & 0.82 & 0.27 & $24.2 \times 28.8$ \phantom{2}(18) & & \\
\rowcolor{Gray}
 & cnt 748 & 10-Mar-16 & 6.0 & $5.0 \times 6\phantom{0}$ & 1.1 & 1.02 -- 1.18 & 0.86 & 0.48 & $24.1 \times 24.9$ (177) & & \\
\rowcolor{Gray}
 & cnt820 & 10-Mar-16 & 6.0 & $5.0 \times 6\phantom{0}$ & 1.1 & 1.02 -- 1.18 & 0.86 & 0.54 & $25.4 \times 26.6$ (172) & & \\
 R\,Crt & NR & 19-Dec-15 & 64.0 & $5.0 \times 16$ & 1.2 & 0.73 -- 0.97 & 0.83 & 0.30 & $ 35.5 \times 38.1$ \phantom{17}(9) & $21.4 \times 23.9$ (4) & 8.5 \\ 
& cnt 748 & 08-Jan-16 & 24.0 & $1.2 \times 20$ & 1.3 & 0.88 -- 1.18 & 0.71 & 0.19 & $37.3 \times 42.4$ (147) & $24 \times 31$ (142) & 6.0 \\ 
& cnt 820 & 08-Jan-16 & 24.0 & $1.2 \times 20$ & 1.3 & 0.88 -- 1.18 & 0.71 & 0.25 & $35.2 \times 40.2$ (153) & $20.1 \times 26.5$ (145) & 4.0 \\ 
\rowcolor{Gray}
HD\,96364 &  NR & 19-Dec-15 & 7.7 & $1.2 \times 16$ & 1.2 & 0.68 -- 0.89 & 0.81 & 0.25 & $ 28.3 \times 29.8$ \phantom{2}(17) & & \\
\rowcolor{Gray}
 & cnt 748 & 08-Jan-16 & 5.6 & $2.0 \times 14 $ & 1.2 & 0.86 -- 0.95 & 0.72 & 0.20 & $28.1 \times 29.5$ (167) & & \\
\rowcolor{Gray}
 & cnt820 & 08-Jan-16 & 5.6 & $2.0 \times 14$ & 1.2 & 0.86 -- 0.95 & 0.72 & 0.26 & $28.5 \times 30.5$ (174) & & \\
\hline                                             
\end{tabular}
\tablefoot{The observations of SW\,Vir for which a neutral density filter (ND1) was used are marked using italic font for the filter name.
DIT is the exposure time of the individual frames, and NDIT is the number of frames per exposure. Each cycle (DIT $\times$ NDIT) was repeated four times
to obtain $+Q$, $-Q$, $+U$, and $-U$ frames, and the whole cycle was repeated several times for each filter and epoch until the total exposure time was reached.
AM and $\theta$ are the airmass and the visible seeing at the time of the observations, and $S_{\rm H}$ and $S_{\lambda}$ are the Strehl ratios observed at the H-band and this ratio calculated at the observed wavelength (see Section~\ref{sec:data}), respectively.
$\lambda_\circ$ and $\Delta \lambda$ are the central wavelength and the FWHM of the filters used. FWHM$_\star$ and FWHM$_{\rm PSF}$ are the
FWHM obtained by fitting a 2D Gaussian to the observed images of the science targets and the PSF references, while
FWHM$^{\rm Mod}_\star$ is the FWHM of the model Gaussian star that best fits the observations.
The position angle of the Gaussian ellipse is given in parentheses after the minor and major values of the FWHM.
Peak PD is the observed peak value of the polarisation degree. $^\dagger$ - For W~Hya, we used the observations of p~Vir as the PSF reference for the fitting procedure (see text).}
\end{table*}

 \begin{figure*}[t]
   \centering
      \includegraphics[width= 18cm]{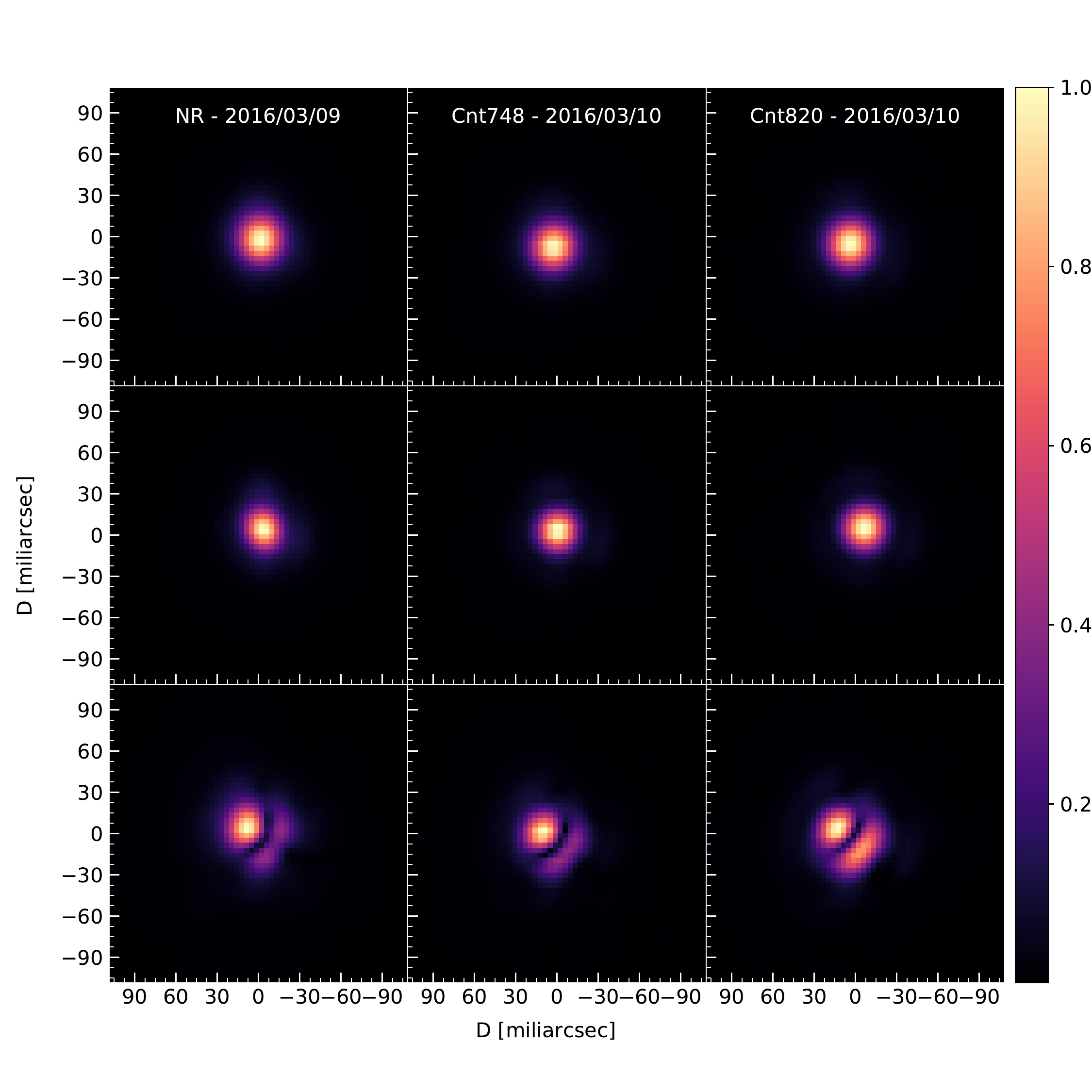}
      \caption{ZIMPOL observations of SW\,Vir in filters NR, cnt748, and cnt820. 
      {Top panels}: Total intensity observed towards SW\,Vir normalised to the peak intensity image.
      {Middle panels}: Images of the PSF-reference star p~Vir normalised to the peak intensity in the image.
      {Bottom panels}: Polarised intensity observed towards SW\,Vir normalised to the peak polarised intensity in the image.
      }
         \label{fig:filtersSWVir}
   \end{figure*}

 \begin{figure*}[t]
   \centering
      \includegraphics[width= 18cm]{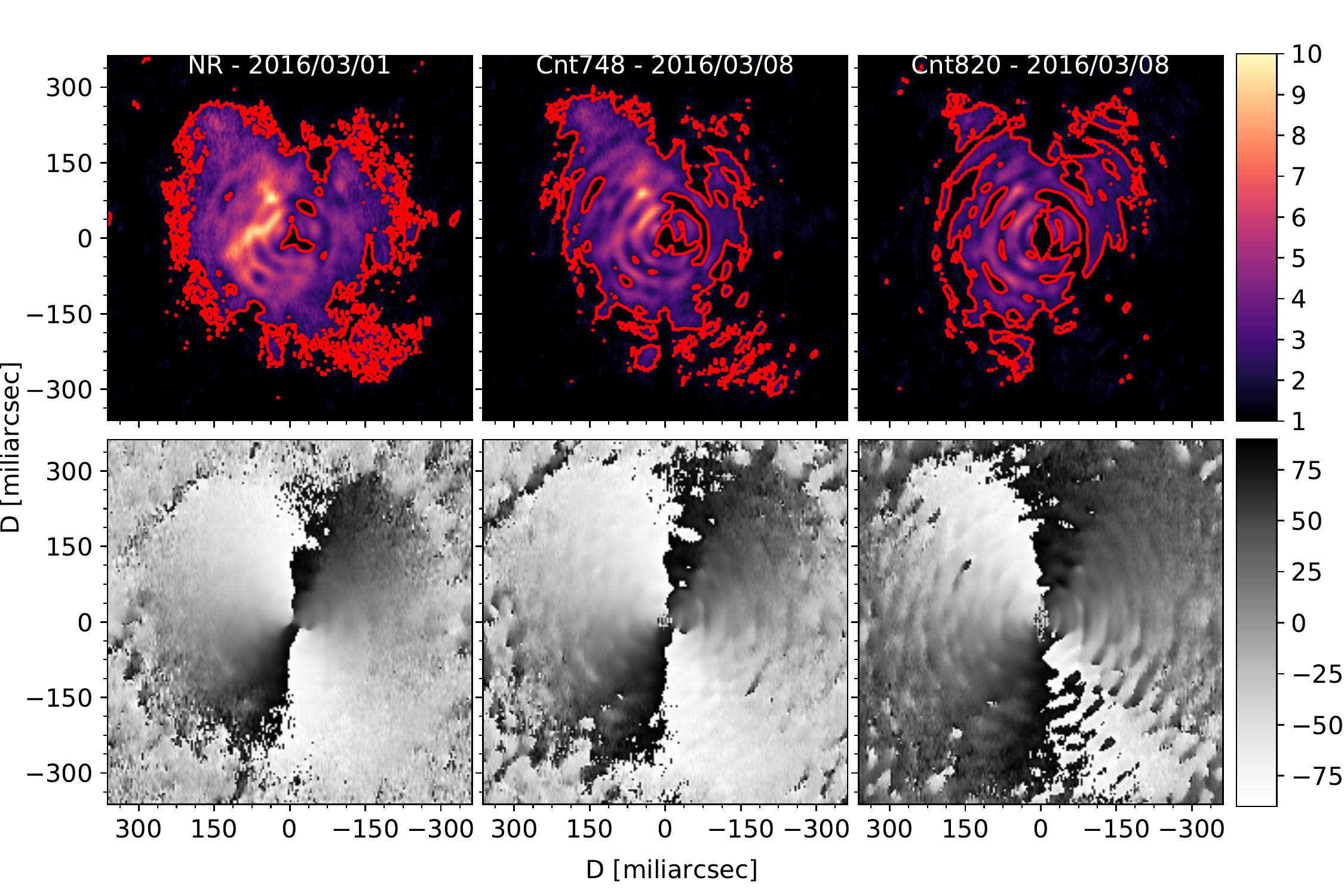}
      \caption{Polarised light in the close environment of SW\,Vir in filters NR, cnt748, and cnt820.
      {Upper panels:} Colour maps showing the percentage polarisation degree observed towards SW\,Vir in filters NR, cnt748, and cnt820 (from left to right).
      The red contours show the 2\% level of the polarisation degree below which instrumental polarisation can dominate.
      {Bottom panels:} Grey-scale maps showing the polarisation angle in degrees with respect to the north direction
       measured in filters NR, cnt748, and cnt820 (from left to right).}
         \label{fig:filtersPolSWVir}
   \end{figure*}

\subsection{Data acquisition and data reduction}
\label{sec:data}

W\,Hya, SW\,Vir, and R\,Crt were observed using ZIMPOL
in three filters: NR ($\lambda_{\rm c} = 0.65~\mu$m), cnt748 ($\lambda_{\rm c}=0.75~\mu$m),
and cnt820 ($\lambda_{\rm c} = 0.82~\mu$m),
with $\lambda_{\rm c}$ the central wavelength of the filter. The filters NR, cnt748, and cnt820 have spectral widths of 56.7, 20.6, and 19.8~nm, respectively.
The ESO ID for the observation programme is 096.D-0930.
R\,Crt was observed in December 2015 and January 2016, while W\,Hya and SW\,Vir were observed in March 2016. The exposure time per source per
filter varied between 15 min and roughly one hour.
SW\,Vir saturated the detector in the observations using filters cnt748 and cnt820. We therefore also acquired
images with the use of a neutral density filter that blocks $\sim 92\%$ of the stellar light (ND1).
Each observing cycle consists of four loops to obtain the images $Q_+$, $Q_-$, $U_+$, and $U_-$, with the half-wave plate oriented at 0$^\circ$,
22.5$^\circ$, 45$^\circ$, and 67.5$^\circ$ with respect to the north direction, respectively. In each loop,
a number of images, NDIT, was acquired with a given detector integration time, DIT. The whole cycle was repeated several times for each filter until
the total exposure time was reached.
Observations of a reference star to obtain the point spread function (PSF) were carried out following the observations of the science target for each filter.
In Table~\ref{tab:obs} we give the exposure and detector integration times for each filter, as well as the
airmass and visible seeing at the time of each observation and the H-band Strehl ratio in the obtained images. The details of the observations of the PSF-reference sources are also given.
The Strehl ratio at the wavelengths of a given ZIMPOL image ($S_\lambda$) was calculated from the H-band Strehl ratio using the Mar\'echal approximation, as done
by \cite{Adam2019} for ZIMPOL observations of IK~Tau following \cite{Lawson2000}.
In this way, $S_\lambda$ is given by
\begin{equation*}
S_\lambda = {\rm exp}\left[\left(\frac{\lambda_{\rm H}}{\lambda}\right)^2 \times {\rm ln}(S_{\rm H})\right],
\end{equation*}
where $\lambda_{\rm H} = 1.65$~$\mu$m. The values obtained are given in Table~\ref{tab:obs} for each observation.

The data were reduced using the (SPHERE/ZIMPOL) SZ-software package developed at the
Eidgen\"ossische Technische Hochschule, Zurich. The basic procedures are standard for astronomical imaging
\citep[bias frame subtraction, cosmic ray removal, and flat fielding; see e.g.][]{Schmid2017}
and equivalent to those carried out using the SPHERE DRH-software provided by ESO.
The images of the Stokes parameters $Q$, $U$, and $I$ were created using
\begin{equation*}
Q = \frac{Q_+ - Q_-}{2},~U = \frac{U_+ - U_-}{2}~~{\rm and}~~I = \frac{Q_+ + Q_- + U_+ + U_-}{4}.
 \end{equation*}
The polarisation observations also require a calibration of the polarimetric modulation-demodulation efficiency of ZIMPOL.
This was done by dividing the observed $Q$ and $U$ images by the polarisation efficiency (typically $\sim 80\%$ for fast polarimetry observations;
SPHERE manual version 101). We performed this correction using the images for the polarisation efficiency calibration provided in the ESO
archive and calculating a mean value of the polarisation efficiency over the image. This was done to minimise noise introduced by image division.
The polarisation intensity, $I_{\rm p}$, polarisation degree, $p$, and direction of the polarisation vectors with respect to the north direction,
$\theta$, were calculated using

\begin{equation*}
I_{\rm p} = \sqrt{Q^2 + U^2},~p = \frac{I_{\rm p}}{I}~~{\rm and}~~\theta = \frac{{\rm arctan}(U/Q)}{2}.
 \end{equation*}

At the time of the observations, the atmospheric wind speeds at the VLT site were always $\gtrapprox 6$~m~s$^{-1}$. This guarantees that the low-wind effect, which can significantly
degrade the obtained PSF, does not affect the observations we present \citep[see][]{Schmid2018}.

The pulsation phase of W\,Hya was 0.5 (minimum light) when the ZIMPOL observations we report were taken.
The low-amplitude and irregular light curves of SW\,Vir and R\,Crt make it difficult
to define their pulsation phases at the time of the ZIMPOL observations.

\section{Observational results}

\subsection{Total intensity images}

The SPHERE/ZIMPOL images resolve the stellar discs of the nearest AGB stars at the $\sim 25$ -- $30\,{\rm mas}$
scale.
In the observations that we report, W\,Hya is clearly resolved, R\,Crt is marginally resolved, and SW\,Vir is not resolved.
We determined full width at half maximum (FWHM) stellar sizes by fitting Gaussian stellar models to the total intensity images
of the three stars using the respective PSF reference images.
This was done by approximating the reference PSF by
a 2D Gaussian and using the {\texttt{imfit}}  routine in the Common Astronomy Software Applications \citep[CASA,][]{McMullin2007} package.
Our results are listed in Table~\ref{tab:obs}.
All the images presented in the paper are the direct result of the reduction process, and no additional deconvolution procedure was applied to the data.

\subsubsection{W\,Hya}
The total intensity (Stokes $I$) images reveal an asymmetric source with significant structure,
with an emission peak in a bean-shaped region in the north-west (see Fig. \ref{fig:filters}).
The morphology of W\,Hya is very similar in filters NR and cnt748, even though the images in these filters were obtained one week apart.
The asymmetry is less prominent, but still obvious, in the image in filter cnt820.
\cite{Ohnaka2017} found very similar morphologies of W\,Hya based on independent
ZIMPOL observations taken about two to three weeks later than those reported here.

The PSF reference for the observations in NR is broader than those in cnt748 and cnt820, probably because of the different conditions when these observations were
carried out.
Moreover, a significantly lower Strehl ratio is retrieved from the images of the PSF-references stars of W~Hya than from those of W~Hya itself.
This suggests that the adaptive-optics system did not perform the same for the observations of the science target and the reference
stars in this case. Although the observations of p~Vir were acquired on different days, they show Strehl ratios 
much more similar to those of W~Hya.
The more comparable Strehl ratios between the images of W~Hya and p~Vir make p~Vir a better suited PSF reference.
This is also confirmed by our more detailed analysis of W~Hya presented in Section~\ref{sec:totInt}.
We therefore used the observations of p~Vir as reference for the PSF in the analysis of W~Hya throughout the paper.
For reference, we infer the FWHM stellar size of W~Hya to be between 5\% and 10\% smaller when considering the PSF-reference stars observed close in time to W~Hya
than when considering p~Vir.

Based on the FWHM stellar sizes obtained using CASA,
we find that the FWHM size of the stellar disc decreases from roughly 65 mas in filter NR to roughly 53 mas in filter cnt820.
This difference is expected to reflect the lower molecular opacity in the wavelength range of filter cnt820.
From our model calculations presented in Section~\ref{sec:totInt}, we derive a uniform stellar disc with a diameter of $\sim 69$~mas by fitting the total intensity images at 0.82~$\mu$m.
Although this value differs significantly from the FWHM derived for a Gaussian stellar disc, this is expected because these two quantities are not directly comparable
because the assumed stellar disc models are different.

\cite{Vlemmings2017} derived the uniform stellar disc size of W~Hya to be $50.9 \times 56.5$~mas in the sub-millimeter at pulsation phase 0.3.
Therefore W~Hya appears larger at visible wavelengths ($> 69$~mas) than at 338~GHz even when the difference in pulsation phase is considered.
\cite{Ireland2004} measured the size of the stellar disc of W\,Hya to be between 0.68 and 0.93~$\mu$m at phase 0.44. The authors reported FWHM sizes of 60~mas at 0.68~$\mu$m, 52~mas
at 0.75~$\mu$m, and 42~mas at 0.82~$\mu$m, using Gaussian stellar discs.
The FWHMs at 0.75~$\mu$m and 0.82~$\mu$m are significantly smaller than what we find.
The filter NR covers the spectral region between 0.617 and 0.674~$\mu$m and therefore does not overlap with the region
observed by \citeauthor{Ireland2004}.
The observations presented by \citeauthor{Ireland2004} and by us where acquired at similar pulsation phases, but the size of W~Hya
at a given pulsation phase seems to vary significantly between pulsation cycles.
We note again that the Gaussian source model
used by \citeauthor{Ireland2004} and us oversimplifies the source morphology, especially for well-resolved observations
as those we present.

\subsubsection{R\,Crt}

The source size is comparable to the resolution of the ZIMPOL images and is therefore marginally spatially resolved. The 
Gaussian stellar disc obtained from the fitting procedure reveals
an elliptical morphology of R\,Crt in the total intensity images in filters cnt748 and cnt820 with a ratio between the short and long axis $\sim 0.76$ and a
position angle of $145^\circ \pm 8^\circ$.
These values are consistent with those obtained based on mid-infrared observations using MIDI of 0.7 at a position angle of $157^\circ$ \citep{Paladini2017}.
The elongation of the stellar disc we report is already evident in the total intensity images presented in Fig.~\ref{fig:filtersRCrt}.
The direction of the elongation is also consistent with that measured by \cite{Ishitsuka2001} for a position angle of 136$^\circ$ of
the biconical outflow traced by H$_2$O masers.
Interestingly, the total intensity image in filter NR does not show an elongated source. The observations in filter NR were acquired 20 days before those in the other two filters.
It is not clear from the data at hand whether this difference is caused by variability of the source, wavelength dependence of the morphology, or
instrumental effects when
some of the observations were acquired.

\subsubsection{SW\,Vir}
\label{sec:totInt_SWVir}

The fitting procedure indicates that the stellar disc of SW\,Vir is unresolved in the obtained images and its FWHM is roughly between 12~mas and 16~mas. This is similar to the
stellar size measured in the near-infrared \citep{Ridgway1982,Schmidtke1986,Mondal2005}
and differs from what we find for W\,Hya and R\,Crt, which appear significantly larger in the visible than in the near-infrared.
The total intensity images of SW\,Vir show strong PSF diffraction rings. This is not apparent in the total intensity images
shown in Fig.~\ref{fig:filtersSWVir}, but can be clearly seen in the images of the polarisation degree (Fig.~\ref{fig:filtersPolSWVir}).
These diffraction rings are very prominent because SW\,Vir is very bright, spatially unresolved in the observations, and was observed in relatively narrow-band filters.
This produces a strong diffraction pattern
that remains prominent because of the small wavelength range probed. As expected, the effect is stronger the narrower the filter used in the observation.
Hence, it is stronger in filters cnt748 and cnt820 than in filter NR.

\subsection{Polarised light}
\label{sec:pol-light}

The ZIMPOL observations show large amounts of polarised light, $I_p$, towards the three sources.
For all stars, the scattered light observations show that dust grains form already at $r \lesssim 2~R_\star$.
The polarisation vectors mostly have tangential directions, as expected for an optically thin envelope irradiated by a central source
(see Figs. \ref{fig:filtersPol}, \ref{fig:filtersPolRCrt}, and \ref{fig:filtersPolSWVir}). In the case of SW\,Vir, this pattern is disrupted by strong PSF
rings (see Section~\ref{sec:totInt_SWVir}).
The polarised light arises from a region consistent with condensation of either aluminium oxide or Fe-free
silicate dust \citep{Hoefner2016}.
Models for dust formation and growth predict
aluminium oxide grains to reach sizes $\gtrsim 0.1~\mu$m, comparable to those of silicates. Aluminium oxide grains might therefore also be expected to
produce polarised visible light.
However, recent results based on observations of gas and dust suggest that aluminium oxide grains
might be smaller than predicted by wind-driving models \citep{Decin2017,Khouri2018}.
Because the scattering properties of the relevant dust species are only weakly dependent on grain composition,
constraining the composition of grains seen through scattered light is not possible from the polarised-light images.
Combining observations using ZIMPOL with high-angular-resolution observations of the gas (e.g. using ALMA) or
dust thermal emission (e.g. using the VLTI) is probably the best way
to constrain the composition of these grains \citep[such as the analysis of Mira based on ALMA and ZIMPOL observations by][]{Khouri2018}.

A comparison between the images of the polarisation degree, $p$, for the three sources (see Figs. \ref{fig:filtersPol}, \ref{fig:filtersPolRCrt}, and \ref{fig:filtersPolSWVir})
reveals that as the wavelength of observation increases, the polarisation degree also increases for W\,Hya, while it decreases for R\,Crt and SW\,Vir.
These different trends are also clear when we consider only the images obtained using filters cnt748 and cnt820 that were
acquired simultaneously for each source. This means that different sky conditions at the times of observations cannot be invoked to explain
the observed difference.
Moreover, the resolution of the images is similar in all filters (but lower in filter NR for our observations) and cannot be invoked either to explain the
decrease in polarisation degree. Therefore we conclude that the different wavelength dependence reflects intrinsic differences between the three sources
at the time of the observations. 
We note that the observed maxima polarisation degrees differ between the three filters at most by $\sim 20\%$, $\sim 35\%$, and $\sim 110\%$ towards W~Hya, SW~Vir, and R~Crt, respectively.

In this type of observations, the performance of the adaptive-optics system is very important in determining how much polarised light is recovered.
The polarised intensity decreases for deteriorating resolution because opposite polarisation vectors
cancel out within a given resolution element. As an example, we convolved images of model envelopes
obtained using the radiative transfer code {\texttt{MCMax}} \citep[see Section~\ref{sec:model},][]{Min2009}
with the PSF-references observed for R\,Crt, SW\,Vir, and W\,Hya in filter NR and find that 
the integrated polarised fluxes measured
after convolution decrease to 50\%, 44\%, and 33\%, respectively, with respect to the original value
of a model with one-pixel ($3.6 \times 3.6$~mas$^2$) resolution.
As mentioned above, the sky conditions became worse between the observations of W\,Hya and those of the PSF-reference sources.
Therefore, the amount of polarised intensity lost is probably smaller in the observations of W~Hya than the original PSF-reference stars.
Nonetheless, this illustrates how the resolution element of the observations (with respect to the size of the emission region)
affect the amount of polarised light retrieved, and the difficulty in directly comparing observations of different sources directly from observed values.

\subsubsection{W\,Hya}

The surface brightness distribution of polarised light measured towards W\,Hya is not symmetric.
Moreover, while the images in filters cnt748 and cnt820 obtained simultaneously are fairly similar, the image in filter
NR obtained one week earlier is somewhat different from those by showing, for instance, peak emission at a difference location.
In the image in filter NR, the polarised intensity peaks in the north-east region, and in those in the cnt748 and cnt820 filters,
it has a broader peak in the north region (see Fig. \ref{fig:filtersPol}). 
Observations by \cite{Ohnaka2017} using ZIMPOL do not show similar differences in morphology. These data were
acquired about two to three weeks later than those we report. This suggests that the difference we find in the brightness distribution
of the polarised intensity between filter NR and filters cnt748 and cnt820 is either a consequence of time variability, or, more likely, caused
by the poorer resolution in the NR image (see Table~\ref{tab:obs}).

We find that the polarised intensity peaks at smaller radii in the image in filter cnt820 than it does in the images in filters cnt748 and NR. This comparison is not straightforward,
however, because defining the central pixel in the three images is not obvious, and for a given direction, the difference is
only about one pixel. These two factors make it difficult to make a comparison as a function of azimuthal angle. Nonetheless, we are able to conclude
that the polarised intensity peaks on average about one pixel (3.6 mas, or 0.35~au for at distance of 98~pc)
farther away from the central star on average in the images in filters NR and cnt748
than it does in the images in filter cnt820. This can be seen from the radial plots in Fig. \ref{fig:radialProf}. As discussed by \cite{Khouri2016}, this might be caused by
increasing scattering optical depths from cnt820 to cnt748 to NR or by the stronger molecular absorption close to the star in filters NR and cnt748.
The polarisation degree reaches peak values of 16.5\%, 18\% and 20\% in filters NR, cnt748, and cnt820, respectively. The three images are shown in Fig. \ref{fig:filtersPol}

\subsubsection{R\,Crt}

The images of the polarised light around R\,Crt reveal a biconical outflow. This is best seen in the images of the polarisation degree shown in Fig.~\ref{fig:filtersPolRCrt}.
The structure is clear in the image using filter NR and is still visible in those using filter cnt748, but not in filter cnt820.
The polarisation degree peaks in the southern half of the biconical outflow and reaches values of 8.5\% in filter NR, 6\% in filter cnt748, and 4\% in filter cnt820.
We estimate the position angle of the biconical structure to be 165$^\circ \pm 15^\circ$ in the images in filter NR and 160$^\circ \pm 10^\circ$ in that in filter cnt748.
These values are higher than those reported by \cite{Ishitsuka2001} for the biconical outflow traced by H$_2$O masers ($\sim 150^\circ$), but given the asymmetric structure and the
consequent large uncertainty in the derived position angles, the values are consistent.

\subsubsection{SW\,Vir}

As noted in Section~\ref{sec:totInt_SWVir}, the PSF diffraction pattern of SW\,Vir strongly affects the images of the polarisation degree
(Fig.~\ref{fig:filtersPolSWVir}).
The PSF diffraction pattern can also be seen in the images of the
polarised intensity (Fig.~\ref{fig:filtersSWVir}), although much less prominently.
This is in principle not expected because the diffraction pattern seen in the two orthogonal polarisation directions should be the same, and should therefore be
erased when the polarised intensity images are created. The explanation for the persisting diffraction pattern might reside in the fact that
ZIMPOL is known to produce a small (sub-pixel) shift between the images of the two orthogonal polarisation directions \citep{Schmid2018}.
This so-called beam-shift effect is well known but difficult to correct for because its strength varies between datasets,
between images in the same dataset, and with telescope pointing position and
rotational states of the optical components. Although the effect can be corrected for by re-aligning the images of the
two polarisation components, we have not ventured into this procedure because the detailed modelling of
the ZIMPOL observations of SW\,Vir is further complicated by the fact that the inner radius of the dust envelope is not well-resolved.
Nonetheless, the observations show that the distribution of dust is not symmetric, with polarised light arising mostly from the eastern hemisphere.

\subsection{Particle sizes}

A general rule is that for small ($2\pi a < \lambda$, where $a$ is the dust particle size)
spherical particles, scattering is close to isotropic and the degree of polarisation is a bell-shaped curve peaking at $90^\circ$ (Rayleigh scattering), while for larger particles ($2\pi a \gtrsim \lambda$) scattering
becomes increasingly forward peaking and the angle-dependence of the degree of polarisation is more complex and varies strongly with particle size.
Moreover, when $2\pi a \sim \lambda$ the scattering properties
depend significantly on the grain models assumed. In the small-particle regime, the scattering cross section decreases with increasing wavelength.

We find that the particles that produce the polarised light around W\,Hya must have radii $\gtrsim 0.1~\mu$m to reproduce the
higher (or comparable) polarisation degree in the images using filter cnt820
than in those using the other two filters. On the other hand, the higher polarisation degree in filter NR relative to those in the other two
filters for R\,Crt and SW\,Vir favours smaller particles, with radii $\lesssim 0.1~\mu$m.
The polarisation degree towards R~Crt drops by more than a factor of two between 0.65~$\mu$m and 0.82~$\mu$m.
We note that although we make this comparison based in general considerations, the exact estimated grain sizes depend on the assumed grain model
and are quite uncertain. We find that the level of polarisation degree observed towards R\,Crt and SW\,Vir is
in agreement with predictions from optically thin models in visible wavelengths using the distribution of hollow spheres \citep[DHS,][]{Min2003} approximation with sizes $< 0.1~\mu$m. We did not
attempt to fit the observed spectral dependence, however.
We note that the polarisation degree can differ significantly from the predictions of spheric models if scattering occurs at a small range of
angles and the grains are not much smaller than the wavelength. This might be
particularly important in the case of R\,Crt, but a better understanding of
the morphology of the source is needed for more detailed model calculations.

Another interesting point is that the size of the particles around W\,Hya has been seen to vary with the pulsation phase of the star \citep{Ohnaka2017}.
It is not clear whether a similar variation occurs for SW\,Vir and R\,Crt.
We note that the light-curve amplitudes of R\,Crt and SW\,Vir are smaller and their periods are shorter than those of W\,Hya.
This means that these stars might present smaller variations in grain properties through their pulsation cycle.
Monitoring of sources with properties similar to those of SW\,Vir and R\,Crt using ZIMPOL will help understand the
mass-loss process in oxygen-rich AGB stars with low pulsation amplitudes and periods.
No theoretical predictions are available for the
light-phase variation of the properties of the dust particles around a star with these characteristics because wind-driving models
have been calculated thus far using pulsation periods $> 300$~days \citep[e.g.][]{Bladh2015,Hoefner2016}. 
Models that consider the formation of dust in outer layers of convective model stars \citep[e.g.][]{Hoefner2019} will
help shed light on this in the future.

\section{Results and discussion}

\subsection{Observed morphology of the envelope of R\,Crt}

There is a clear difference between the biconical morphology of the envelope of R\,Crt and the more spherical (but still asymmetrical)
envelopes of W\,Hya and SW\,Vir.
Because R\,Crt has the highest mass-loss rate of the 43 semi-regular variable AGB stars observed by \cite{Olofsson2002},
the observed morphology can be an indication that either the higher mass-loss rate is produced by a
peculiar process that also produces the biconical outflow, or that the inferred mass-loss rate from CO lines using
a spherically symmetric model is not accurate for this source.
We note that it is not clear, however, whether the biconical morphology of the
circumstellar envelope close to the star also extends
to larger distances where the CO lines are mainly excited.
The lower mass-loss rate derived from the OH maser observations \citep[$\sim 1 \times 10^{-7}~\dot{M}~{\rm yr}^{-1}$,][]{Szymczak1999}
is also interesting in this context but does not help solve the problem because the derived OH shell
size also depends on the assumption of a spherically symmetric and thin OH shell.
The position of the SiO ring at a distance of roughly twice the near-infrared stellar radius
is typical for AGB stars \citep[e.g.][]{Diamond1994,Cotton2004}, and the observed maser spots do not
trace a biconical shape. H$_2$O maser emission is only prominent in the south-east cone \citep{Kim2018},
which also appears brighter in polarised visible light. The unknown inclination angle precludes us from
concluding whether this is caused by an asymmetry in the mass distribution between the two cones
because of possible differences between forward and backward scattering.
The southwest cone seems to be inclined towards the observer because H$_2$O maser spots are blue-shifted with respect to the velocity of R\,Crt.

A biconical or bipolar outflow is not common around AGB stars but is observed in a few sources, including V~Hya~\citep{Sahai2003},
CIT~6~\citep{kim2015}, and WX~Psc~\citep{Vinkovic2004}.
Interestingly, the mass-loss rates of these stars are all at least one order of magnitude higher than the rate derived for R\,Crt.
The morphology is sometimes associated with the presence of a fast outflow \citep[e.g.][]{Sahai2003}
and it is often taken as an indication of a transition into the post-AGB phase \citep[e.g.][]{Vinkovic2004}.
We note that there are no indications of the presence of a fast outflow in R\,Crt.
The origin of this biconical outflow around these sources is unclear, but shaping caused by a binary companion and/or
by a strong magnetic field are possible scenarios \citep[e.g.][]{DeMarco2017}.
The current theory requires an additional source of angular momentum
to explain magnetic fields in AGB stars \citep[e.g.][]{Nordhaus2007}, which might be provided by a binary companion
or even a planetary system. It is not yet clear whether the ubiquitous occurrence of magnetic fields
in AGB stars can be explained solely by interactions with stellar companions \citep{Vlemmings2019}. 

 \begin{figure*}[t]
   \centering
      \includegraphics[width= 18cm]{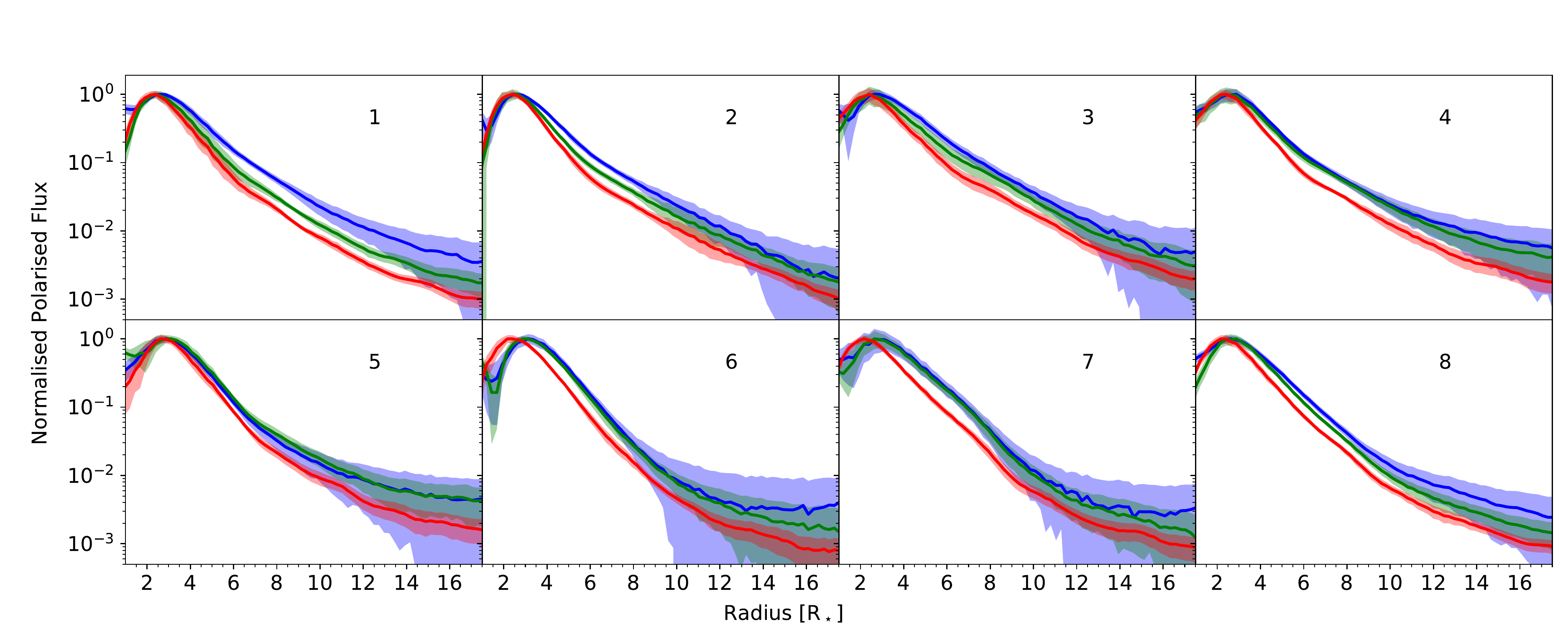}
      \caption{Normalised radial profiles of the polarised intensity observed towards W\,Hya
      for the eight octants of the images using filters NR (blue), cnt748 (green), and cnt820 (red).
      The stellar radius assumed in this plot, $R_\star$, is the near-infrared stellar radius measured by \cite{Woodruff2009} of 18~mas.}
         \label{fig:radialProf}
   \end{figure*}   

Another source that shows a biconical outflow is L$_2$~Pup. ZIMPOL observations of this star
show an edge-on disc obscuring the star in the equatorial direction but
allowing visible light to escape in the polar directions. The disc might also collimate the outflow into a biconical structure \citep{Kervella2015}. The morphology of
the dusty envelope of R\,Crt could be explained by a similar scenario to that of L$_2$~Pup.
An obvious difference between the two sources is
the expansion velocities of the outflows inferred from single-dish observations of CO lines,
of 2.5~km~s$^{-1}$ for L$_2$~Pup and of 10~km~s$^{-1}$ for R\,Crt \citep{Olofsson2002}.
Interestingly, observations with ALMA reveal gas expansion velocities up to $\sim 17$~km~s$^{-1}$,
which is not evident in single-dish spectra \citep{Homan2017}. Spatially resolved observations with higher sensitivity
of molecular line emission towards R~Crt will allow for a better comparison between these two sources.

The mass-loss rates derived based on low-excitation CO lines or dust excess may not be accurate because
they are based on spherically symmetric models.
If a disc that blocks stellar light, and potentially the outflow, exists in the equatorial direction around R\,Crt,
it must be oriented nearly edge on and the biconical structure observed in scattered light must be perpendicular to it. We note, however, that no direct
evidence of such a disc is seen in the ZIMPOL images we present. Moreover, \cite{Paladini2017} reported that the source has a position angle of 157$^\circ$ at $\sim 10~\mu$m, consistent with that of the biconical outflow we find.
If a compact, dense disc exists around R\,Crt, we would have expected its signature to be seen
as excess thermal dust emission in a perpendicular direction from that of the biconical outflow.
ALMA observations of L$_2$~Pup have revealed a disc with Keplerian rotation around the star and allowed the mass of the
possible planetary-mass secondary body to be constrained \citep{Kervella2016}. Equivalent observations of R\,Crt will help
determine whether these two sources have similar morphologies of their circumstellar envelopes.
Regarding the maser emission and the magnetic field orientation, a direct comparison to L$_2$~Pup is not possible because no spatially resolved observations
of either the maser emission region or the magnetic field structure have been reported for L$_2$~Pup to our knowledge.

In conclusion, we are not able to determine the origin of the biconical structure seen in the circumstellar envelope of R\,Crt. A comparison to L$_2$~Pup
suggests that these two sources might have a similar morphology, but no sign of a disc around R\,Crt is seen in either the ZIMPOL or the MIDI data.
An alternative explanation, supported
by the detection of a strong magnetic field in the inner part of the biconical outflow, is that magnetic fields in R\,Crt collimate the outflow.
Further observations aimed at detecting a possible disc around R\,Crt and the maser morphology and magnetic field strength in L$_2$~Pup will help
discern between these two possibilities.

 \begin{figure}[t]
   \centering
      \includegraphics[width= 8cm]{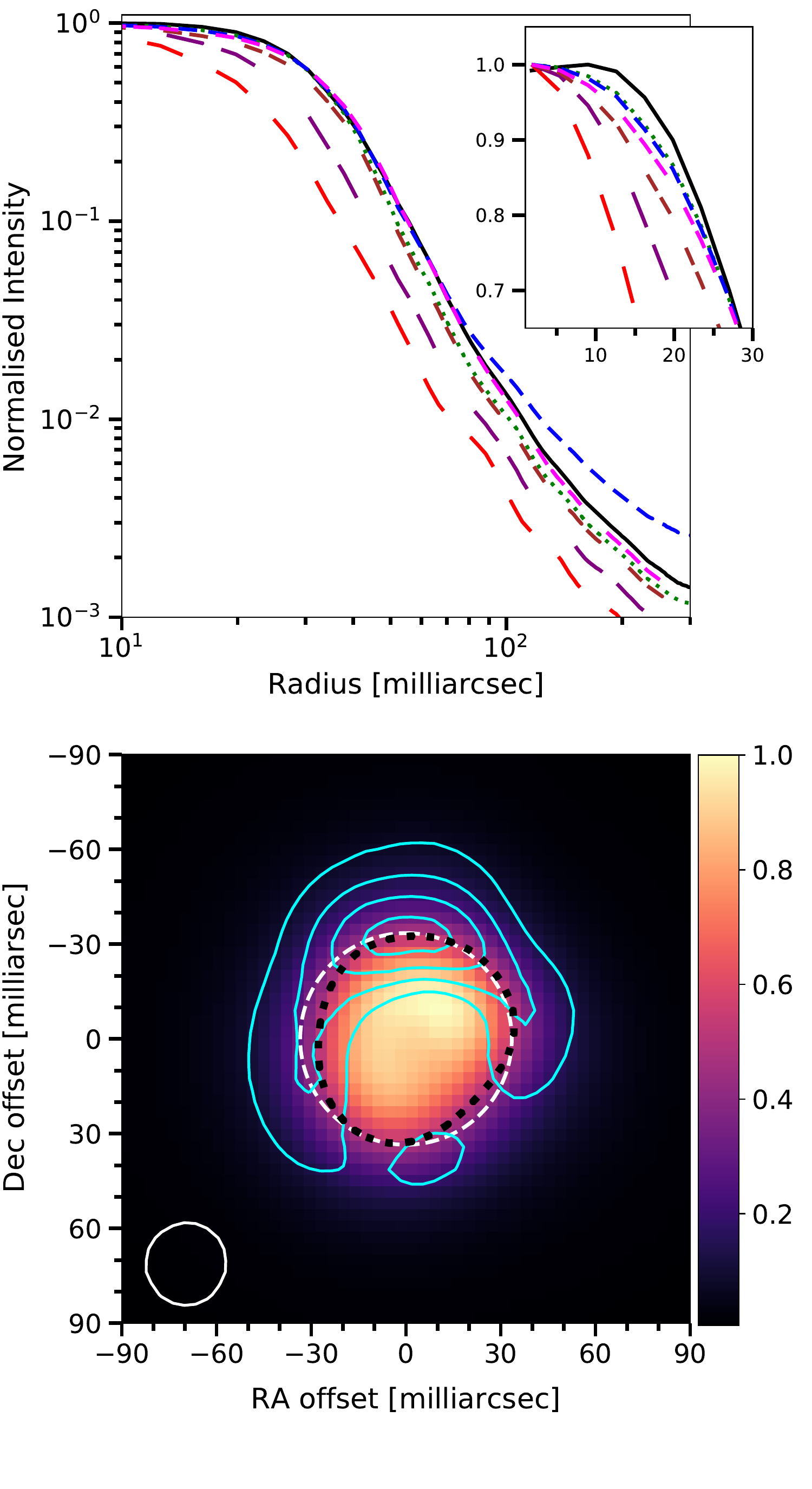}
      \caption{Comparison between models and observations of the total intensity images in the cnt820 filter. {\it Upper panel:} Observed radial profile of the total intensity (full black line)
      compared to radial profiles derived from models convolved with the PSF-reference p~Vir and considering uniform stellar discs with radii 1.86~au (long-dashed red line),
      2.56~au (double-long-dashed purple line), 3.35~au (dotted green line), and 1.86~au surrounded by a dust envelope that is optically thick in the visible due to scattering extending out to 3.35~au
      (double-short-dashed brown line). The short-dashed blue line shows the radial profile obtained from a model with a uniform stellar disc with a radius of 3.35~au convolved with the PSF-reference
      HD~118877, observed following the observations of W~Hya. {\it Lower panel:} Comparison between the observed normalised total intensity (colour-scale image), the polarised intensity
      contours at 90\%, 70\%, 50\%, and 30\% times the peak value
      (full cyan line), and the stellar disc with a radius of 3.35~au (dashed white line). The 50\% levels of the images of the PSF-reference p~Vir (white solid line) and
      of W~Hya (dotted black line) are shown for reference.}
      \label{fig:TotIntRadialProf}
   \end{figure}

\subsection{Modelling of the observations of W~Hya}
\label{sec:model}

We have calculated models to fit the observations of W~Hya. The main goal was to use the polarised-intensity images to derive the radial profile of the
density distribution of the grains that produce the scattered light.
We did not attempt to model the polarised light observed towards R\,Crt and SW\,Vir. For R\,Crt, this is because of the biconical morphology of the envelope.
For SW\,Vir, this is both because the inner radius of the envelope is not well resolved and because of the strong diffraction rings observed
also in polarised light, which would complicate the modelling.

For our models of W\,Hya, we focused on the data obtained using filter cnt820 because molecular and dust scattering opacities
are expected to be lower in this wavelength range than in those of filters NR and cnt748 \citep[e.g.][]{Khouri2016}.
Moreover, together with the images in filter cnt748,
the images in cnt820 show polarised light up to larger distances from the star. In Fig.~\ref{fig:radialProf} we show how the radial profile of the
polarised intensity differs between observations using the three filters.

\subsubsection{Total intensity image}
\label{sec:totInt}

We first focused on reproducing the observed size of the stellar disc of W~Hya. Our model consists of a black-body, uniform-disc star and therefore does not provide a very accurate description
of the observed images, which reveal significant structure (see Figs.~\ref{fig:filters} and \ref{fig:TotIntRadialProf}). Nonetheless, the size (and shape) of the stellar disc significantly affects not only the model
unpolarised-light images, but also the polarised-light and the polarisation-degree ones. Obtaining a model as accurate as possible for the stellar disc is therefore important for our analysis.

By comparing uniform-disc models with different radii to the azimuthally averaged total intensity images in the cnt820 filter, we find that models with radii of $\sim 3.45$~au are required (see Fig. \ref{fig:TotIntRadialProf}). This is much larger than the 1.8~au size of our reference stellar radius. This
dramatic increase between infrared and visible wavelengths is caused by the thick atmosphere of molecules (and dust, as discussed below)
around W~Hya. As shown in the inset in Fig.~\ref{fig:TotIntRadialProf}, our model is unable to reproduce the off-centre peak of the total intensity images.
This is an expected shortcoming of our model, given the simplicity of our model star.

Before comparing the models to the observations, the images were convolved with an image of a PSF-reference star.
We  find that the images of HD~118877 do not provide a good description of the PSF at the time of the observations of W~Hya. This is most likely because of changing sky conditions
between the observations of the two stars, and is evident from the different Strehl ratios obtained for the two sets of observations (see Table~\ref{tab:obs}).
The effect of this is clearly seen in Fig.~\ref{fig:TotIntRadialProf} as an 
excess at large radii when comparing the 3.45~au model convolved with the images of HD~118877 (blue dashed line) and the observations of W~Hya (solid black line). The 
images of p~Vir (originally observed as a PSF-reference for the observations of SW~Vir) were acquired at similar sky conditions to those of W~Hya and therefore
present a more comparable Strehl ratio. We find that using p~Vir as a PSF-reference significantly improves our fits to the data at radii $\gtrsim 80$~mas from the central star. Therefore, we used p~Vir as the
PSF-reference star in
our calculations instead of HD~118877.

We find that the outflowing dust does not significantly affect the total intensity images and the derived stellar size accordingly does not depend strongly on
the outflowing-dust model (especially given the simplicity of our stellar model). In this sense, there is a clear distinction in our model between dust grains residing in the high-density, gravitationally bound region
(which affect the derived stellar size in the visible) and the outflowing dust. In reality, the
transitions between these two regions is much more complex than our model assumes, and more detailed modelling of this transition region is necessary to better constrain the distribution of dust in this region and the effect of scattering on the
stellar size in the visible.

\subsubsection{Polarised intensity image}
\label{sec:polInt_WHya}

We calculated continuum radiative transfer models using the code {\texttt{MCMax}} \citep{Min2009}.
{\texttt{MCMax}}  treats scattering and polarisation of light using the full angle-dependent Mueller matrix
and produces as output images of the $Q$ and $U$ parameters. We convolved these images with the image
of the PSF-reference star adopted by us (p~Vir) before computing the polarisation intensity and the polarisation degree images,
which were then compared to the observations.

We assumed a black-body star surrounded by Mg$_2$SiO$_4$ grains. We used the optical constants provided by \cite{Jager2003} to calculate the opacities and scattering matrix based on
the DHS approximation \citep{Min2003}. Our results are independent of the choice of model star because we do not fit absolute quantities, such as flux density or absolute polarised intensity, and
focus only on relative quantities, such as the normalised radial profiles of the total and polarised intensities and the polarisation degree.
We do not expect the choice of optical constants to significantly affect the results because the other reasonable choices of dust species (MgSiO$_3$ and Al$_2$O$_3$)
have scattering properties that are similar to those of Mg$_2$SiO$_4$.
We chose the DHS grain model (with an irregularity parameter $f_{\rm max} = 0.8$) because it has been shown to produce a good fit
to the scattering properties of real particles measured in the laboratory \citep{Min2005}.
The model dust envelopes we calculated are spherically symmetric and irradiated by a central  star.
   
\subsubsection{Radial profile of the dust density}
\label{sec:radialProf}

Our modelling approach is similar to that
employed by \cite{Khouri2016} for modelling ZIMPOL observations of R~Dor. We divided the observed images into octants and fit the azimuthal average of the
observed radial profiles of the polarised light for each octant. We numbered the octants from one to eight starting with the octant delimited by the north and north-west direction and proceeding clockwise.
The uncertainties in the observed radial profiles are given by the combination of the uncertainty on the polarised intensity per pixel (given by the ESO
pipeline) and of the standard deviation of the polarised intensity between pixels at a same given radial distance.
We ranked the models based on the $\chi^2$ of the fits to the data. The uncertainties on the derived parameter values were calculated by considering all models
that produce $\chi^2$ values that are lower than ${\chi^2_{\rm best} + \sqrt{2\nu}}$, where $\chi^2_{\rm best}$ is the lowest $\chi^2$ value obtained and
$\nu$ is the number of degrees of freedom in the $\chi^2$ calculation.

We find that the choice of grain size (within the range we explored, $0.1~\mu$m to $0.5~\mu$m) and the grain composition (Mg$_2$SiO$_4$, MgSiO$_3$, or Al$_2$O$_3$)
has a much weaker effect than the dust density profile on the radial profile of the polarised intensity. For simplicity, we therefore modelled the radial profiles using
only Mg$_2$SiO$_4$ grains with sizes of 0.1~$\mu$m, as determined by \cite{Ohnaka2017}.

\begin{figure*}[t]
   \centering
      \includegraphics[width= 18cm]{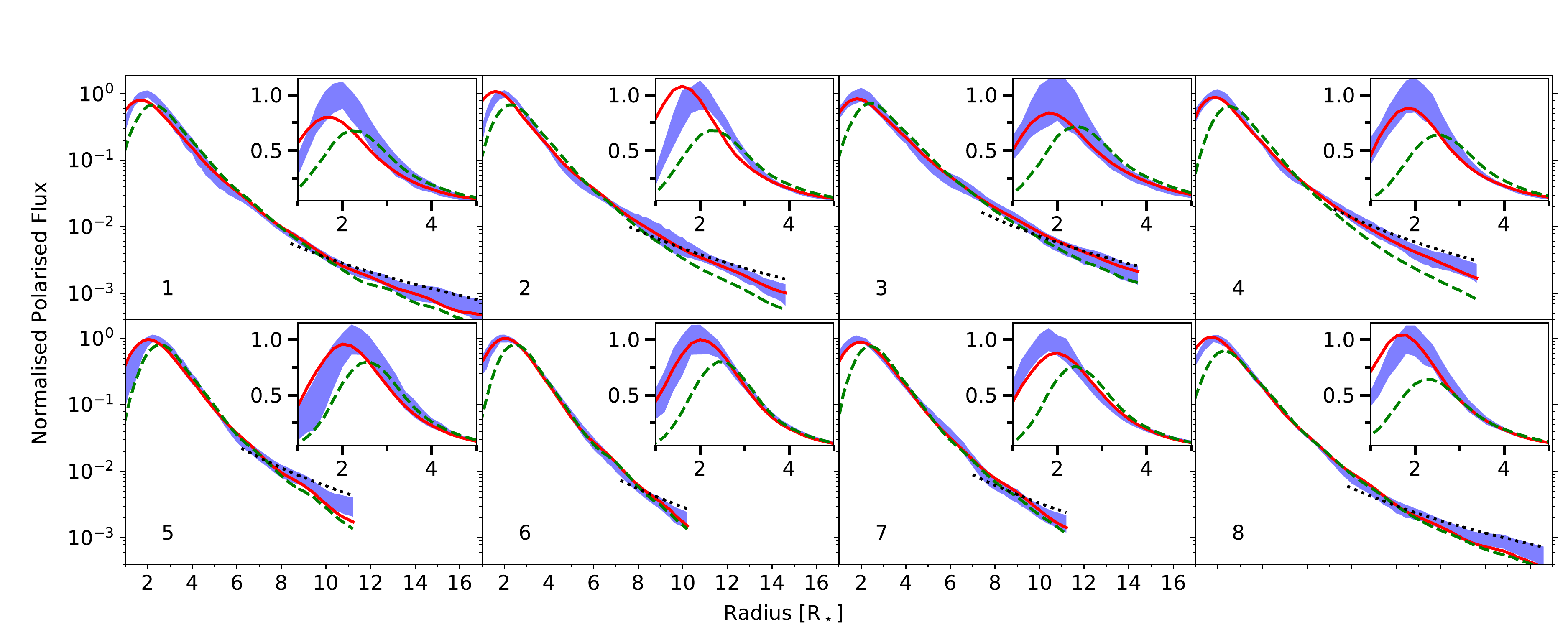}
      \caption{Models to fit the polarised light observed towards W\,Hya for each of the eight octants of the image in filter cnt820.
      The observations with the 1$\sigma$ error bars are shown by the shaded blue region,
      the full red lines show the best-fit models for each octant, the dashed green line shows the best model with a 3.45~au star,
      and the dotted black lines show the best-fit models with a radial power-law index $n=2$.
      The stellar radius assumed in this plot, $R_\star$, is the near-infrared stellar radius measured by \cite{Woodruff2009} of 18~mas.}
         \label{fig:modRadialProf}
   \end{figure*}

The radial dust-density profile of the outflow in our models is described by the power law
$\rho(r) = \rho_\circ\, (R_\circ/r)^{-n}$, where $\rho_\circ$ is the dust density at the inner radius of the dust outflow $R_\circ$.
We are unable to reproduce the observed peak of the radial profile of the polarised intensity with a model with a radius of 3.45~au, which is required to fit the total intensity images.
This is because the minimum possible value of $R_\circ$ (3.45~au) is too high. The only way we could improve the fit to the observations in the context of our model was to substitute
the central source by a smaller star surrounded by an optically thick scattering dust shell with an outer radius of 3.45~au. Such a shell produces sufficient polarised light to shift the peak
of the polarised intensity to smaller radii, providing a much better fit to the observations (see Fig.~\ref{fig:modRadialProf}). For all the octants, the models with a star surrounded by an optically thick shell provide
$\chi^2$ values lower by more than a factor of two than the models with a 3.45~au star and no optically thick dust shell. By comparing the brightness distributions of the polarised and unpolarised
intensity (as shown in Fig.~\ref{fig:TotIntRadialProf}), it is evident that indeed the polarised flux peaks very close to or even within the photosphere at 0.82~$\mu$m.

The large size of the central source implies that $R_\circ$ is effectively a fixed parameter in our fitting procedure because it is always set to 3.45~au. Therefore,
the only free parameters are the exponent $n$, the normalisation of the radial profile of the polarised intensity, and the ratio between the dust density
immediately inside and outside the optically thick dust shell. Our models are insensitive to the inner radius and dust density in the optically thick inner shell,
provided that the shell is optically thick. Considering 0.1~$\mu$m grains, our models require dust densities in the optically thick dust shell of $\gtrsim 2 \times 10^{-16} {\rm g~cm^3}$
and a dust density decrease by a factor between 30 and 100 at the outer edge of the optically thick shell. To place better constraints on the dust distribution in this region, a model that includes
the contribution of absorption and scattering by molecules is necessary.

The resulting best-fitting models are shown in Fig.~\ref{fig:modRadialProf}, and the interpolated best-fitting values for the parameters and the uncertainties are given
in Table~\ref{tab:octRad}. 
Our fitting results suggest that $n$ varies significantly between the different octants, ${3.0 \lessapprox n \lessapprox 4.8}$.
The asymmetry of the dust envelope can be inferred directly from Fig.~\ref{fig:filtersPol} by noting that the polarisation degree drops below 2\% at shorter distances to the east than to the west, even if
the polarisation degree close to the star is higher to the east than to the west.
To reproduce the observed polarisation degree, the outwards radial optical depths required by our models outside of the optically thick shell vary between $\sim 0.5$ and $\sim 1$ for the octants that display minimum and maximum polarised light.

The derived steepness of the dust density profile depends not only on the acceleration profile of the outflow, but also on the contrast between the
amount of
grains in the high-gas-density extended atmosphere (which is gravitationally bound) and the number of grains in the accelerating wind (where radiation pressure has overcome gravity).
Determining the density and velocity distribution of the gas in these inner regions at the time
of ZIMPOL observations will allow for a much more complete picture of the acceleration region of the given AGB outflow.
The dynamical timescale for the dust grains at these distances is about a few years (assuming $\upsilon_{\rm dust} = 15~$km/s) and the timescale for variation of
the dust distribution in the innermost regions seems to be about a month. Therefore, observations of the time variability
of the dust density distribution around W\,Hya over a few months are expected to provide important additional constraints.

Our results also imply that dust might form already inside the visible photosphere at radii $< 1.9~R_\star$. This would not be at odds with theoretical models \citep[e.g.][]{Hoefner2016}, but implies
that inner radii of scattering dust envelopes derived from this type of observations should be viewed as upper limits. The high values of the dust density required by our models of the optically thick
dust shell suggest that at least part of the opacity is due to molecular absorption or scattering. Whether the observed scattered light can be
partially explained by scattering of light by molecules in the high-gas-density regions close to the W~Hya \citep[$\sim 10^{11}~{\rm mol~cm}^{-3}$][]{Vlemmings2017}
needs to be investigated in future studies.

We also considered the possibility that the peak of the polarised intensity is shifted to smaller radii because of instrumental effects that produce spurious polarised light.
One known effect that can significantly affect the polarised signal close to the star in ZIMPOL observations is the beam shift \citep{Schmid2018}. We find no evidence for
significant beam-shift effect in the observations of W~Hya, therefore we consider this an unlikely cause for the shift in polarised-intensity peak to smaller radii in the cnt820 images.

Multi-wavelength imaging or interferometry
observations provide crucial information for the interpretation of our results. For instance, \cite{Ireland2004} observed W~Hya at a pulsation phase 0.44
with the Masked Aperture-Plane Interference Telescope in the Anglo-Australian Telescope to determine the size of the stellar disc of W~Hya at wavelengths between 680 and 930~nm.
The authors found that the stellar size increased smoothly towards the blue, instead of abruptly within molecular bands, and suggested that the observed
photosphere of W~Hya at visible wavelengths must be significantly affected by dust scattering. We note that our model with an optically thick dust shell is unable to
reproduce the increase in stellar size between 0.82 and 0.75~$\mu$m.
However, obtaining a fit with our spherically symmetric model to the images using the cnt748 filter is difficult
because of the higher contrast between the brightness at the centre of the stellar disc and the off-centre emission peak.
Moreover, the importance of dust grains for the total opacity and for determining the size of the stellar photosphere is expected to be relatively larger in a spectral region without strong
molecular features, such as the region covered by the cnt820 filter employed by us. Molecules are expected to have a stronger effect on the appearance of the photosphere
at 0.65~$\mu$m and 0.75~$\mu$m.

\cite{Ohnaka2017} also modelled ZIMPOL observations of W~Hya similar to those presented by us and were able to reproduce the position of the peak of the polarised
intensity. However, the authors did not reproduce the size of the stellar disc and considered a star with a radius of $\sim 1.8$~au,
which our results show might lead to inconsistencies. The inner radius of the dust envelope derived by them (2.3~au) is smaller than the size of the stellar photosphere
at 0.82~$\mu$m determined by us.

\renewcommand{\arraystretch}{1.0} 
\begin{table}
\caption{Radial distances of the regions compared to the models for each octant in the cnt820 images. We also present the relative peak intensity
at each octant (RPI) relatively to the peak intensity of the azimuthal average of the entire image. $R^{\rm exp2.0}_{\rm min}$ is the radius up
to which the radial dust density profile is inconsistent with a non-accelerating steady-state wind.}
\label{tab:octRad}      
\centering                                      
\begin{tabular}{c | c c c c c c} 
Oct & $r$ & RPI & $n$ & $\chi^2_{\rm min}$ & $\nu$ & $R^{\rm exp2.0}_{\rm min}$ \\
 & [mas] & & & & & [mas] \\
\hline
1 & 280 & 1.7 & $3.40 \pm 0.3$ & 37.3 & 70 &137 \\ 
2 & 250 & 1.0 & $3.35 \pm 0.2$ & 63.7 & 58 & 122 \\
3 & 230 & 0.7 & $3.00 \pm 0.4$ & 10.9 & 57 & 119 \\
4 & 230 & 0.6 & $3.00 \pm 0.3$ & 22.0 & 53 & 115 \\
5 & 180 & 0.6 & $4.00 \pm 0.4$ & 26.5 & 44 & 97 \\
6 & 180 & 0.9 & $4.80 \pm 0.3$ & 17.6 & 36 & 115 \\
7 & 180 & 1.0  & $4.10 \pm 0.2 $ & 30.0 & 41 & 112 \\
8 & 280 & 1.7 & $3.85 \pm 0.2$ & 55.7 & 68 & 130 \\
\end{tabular}
\end{table}

\subsubsection{Extent of the acceleration region around W\,Hya}

Independently of the best power-law fit, we also investigated the minimum
radius from which a ${\rho(r) \propto r^{-2}}$ density law is consistent with the data, $R^{\rm exp2.0}_{\rm min}$.
This was done by calculating the cumulative $\chi^2$ starting from the outer edge of each octant and proceeding inwards in radius for a model with $n=2$.
The outer edge was defined as the radius at which the polarisation degree dropped below 2\% (see Table~\ref{tab:octRad}), while
$R^{\rm exp2.0}_{\rm min}$ was defined as the radius at which the $\chi^2$ value deviated more than 3$\sigma$ from the expected value given
the number of points considered and the associated errors.
The motivation for this calculation is that the dust density radial profile of a steady-state outflow
should transition from the steep profile observed close to the star ($n>2$) to a profile with $n=2$ when the wind reaches terminal velocity.
In this way, $R^{\rm exp2.0}_{\rm min}$ is a lower limit to the outer radius of the dust-acceleration region.

Our results show that the $\rho(r) \propto r^{-2}$ density law expected for an outflow expanding at constant velocity is not consistent with the observations for $r \lesssim 100$~mas. Therefore the acceleration of the wind is expected to happen at least up to a radius of $\sim 100$~mas, which is expected.
These are lower limits for the radius from which the dust particles are no longer accelerated.
For the direction in which polarised light is seen up to larger distances, we
find that density law with $n=2$ is not consistent with the observations up to $r \approx 137$~mas. When we consider the stellar radius in the infrared
of 18~mas \citep{Woodruff2009}, this distance
corresponds to $r \approx 7.5~R_\star$. These lower limits are still consistent with theoretical predictions for the wind-acceleration region
\citep[e.g.][]{Hofner2008}, which predict an acceleration region extending to $\sim12~R_\star$.
This shows that high-angular-resolution observations of polarised light are very close to directly testing model predictions for
the wind acceleration in AGB stars.

\section{Summary and conclusions}

We presented ZIMPOL observations of W\,Hya, R\,Crt, and SW\,Vir, 
which show mass-loss rates between 10$^{-7}$ and 10$^{-6}$~M$_\odot$/yr and relatively small
visual light amplitudes (between 1.4 and 2.8 mag).
The images in polarised light show very different morphologies
of the inner dust envelopes.
Particularly, R\,Crt shows a biconical circumstellar envelope which had been
previously seen in observations of maser emission. The cause for
this peculiar morphology is not clear, but we speculate that it might be caused by
interactions with a binary companion and/or strong magnetic fields. This object might provide
important insight into how the shaping of biconical outflows takes place.

The polarisation degree observed toward R~Crt decreases sharply between 0.65~$\mu$m and 0.82~$\mu$m,
while for SW~Vir and W~Hya the observed values are more similar. In the case of W~Hya, the polarisation degree
is maximum at 0.82~$\mu$m. \cite{Ohnaka2017} modelled observations towards W~Hya at a similar phase and
derived grains sizes of 0.1~$\mu$m. The observations we report suggest that the grains around SW\,Vir are $\lesssim 0.1$~$\mu$m
and those around R~Crt are smaller $< 0.1~\mu$m than those around W\,Hya. This indication must be confirmed using radiative
transfer modelling.
It is not clear whether this reflects a fundamental difference because the size of the grains around W\,Hya
has been seen to vary from $\sim 0.1~\mu$m to $\sim 0.5~\mu$m from minimum to maximum light phase \citep{Ohnaka2017}.
The more regular and larger amplitude pulsations of W\,Hya in comparison to R\,Crt and SW\,Vir
might cause the properties of the dust grains to vary more strongly. Monitoring observations of low-amplitude pulsators
will help shed light on this issue.

The stronger polarised signal is detected towards W\,Hya, and we are able to see scattered light out to $\sim 17$ infrared
stellar radii. Our observations show that a significant amount of polarised light is produced within the visible photosphere.
We are able to reproduce this by considering models with an inner dust shell that is optically thick to scattering.
This suggests that inner radii derived from this type of observations should only be regarded as upper limits.
Moreover, we find that the density of dust grains drops steeply outside the photosphere, most likely tracing
the transition between extended atmosphere and accelerating outflow. We are able to determine lower limits to the extent of the acceleration region
and find that acceleration occurs until at least 7 infrared stellar
radii. This is consistent with predictions from theoretical models, which find that outflows accelerate up to $\sim 12$ stellar radii. Our findings highlight the power of ZIMPOL in imaging dust in the crucial
dust-formation and wind-acceleration region around AGB stars.

\begin{acknowledgements}
We thank the anonymous referee whose careful reading of the manuscript and thoughtful suggestions significantly contributed to
improve the quality of the paper.
This work was supported by ERC consolidator grant 	.
EDB acknowledges financial support from the Swedish National Space Agency.
\end{acknowledgements}

\bibliographystyle{aa}
\bibliography{../bibliography_2}

\begin{thebibliography}{74}
\expandafter\ifx\csname natexlab\endcsname\relax\def\natexlab#1{#1}\fi

\bibitem[{{Adam} \& {Ohnaka}(2019)}]{Adam2019}
{Adam}, C. \& {Ohnaka}, K. 2019, \aap, 628, A132

\bibitem[{{Bladh} \& {H{\"o}fner}(2012)}]{Bladh2012}
{Bladh}, S. \& {H{\"o}fner}, S. 2012, \aap, 546, A76

\bibitem[{{Bladh} {et~al.}(2015){Bladh}, {H{\"o}fner}, {Aringer}, \&
  {Eriksson}}]{Bladh2015}
{Bladh}, S., {H{\"o}fner}, S., {Aringer}, B., \& {Eriksson}, K. 2015, \aap,
  575, A105

\bibitem[{{Bladh} {et~al.}(2017){Bladh}, {Paladini}, {H{\"o}fner}, \&
  {Aringer}}]{Bladh2017}
{Bladh}, S., {Paladini}, C., {H{\"o}fner}, S., \& {Aringer}, B. 2017, \aap,
  607, A27

\bibitem[{{Bowers} \& {Johnston}(1994)}]{Bowers1994}
{Bowers}, P.~F. \& {Johnston}, K.~J. 1994, \apjs, 92, 189

\bibitem[{{Cotton} {et~al.}(2004){Cotton}, {Mennesson}, {Diamond}, {Perrin},
  {Coud{\'e} du Foresto}, {Chagnon}, {van Langevelde}, {Ridgway}, {Waters},
  {Vlemmings}, {Morel}, {Traub}, {Carleton}, \& {Lacasse}}]{Cotton2004}
{Cotton}, W.~D., {Mennesson}, B., {Diamond}, P.~J., {et~al.} 2004, \aap, 414,
  275

\bibitem[{{Cox} {et~al.}(2012){Cox}, {Kerschbaum}, {van Marle}, {Decin},
  {Ladjal}, {Mayer}, {Groenewegen}, {van Eck}, {Royer}, {Ottensamer}, {Ueta},
  {Jorissen}, {Mecina}, {Meliani}, {Luntzer}, {Blommaert}, {Posch},
  {Vandenbussche}, \& {Waelkens}}]{Cox2012}
{Cox}, N.~L.~J., {Kerschbaum}, F., {van Marle}, A.-J., {et~al.} 2012, \aap,
  537, A35

\bibitem[{{Danilovich} {et~al.}(2017){Danilovich}, {Lombaert}, {Decin},
  {Karakas}, {Maercker}, \& {Olofsson}}]{Danilovich2017}
{Danilovich}, T., {Lombaert}, R., {Decin}, L., {et~al.} 2017, \aap, 602, A14

\bibitem[{{De Marco} \& {Izzard}(2017)}]{DeMarco2017}
{De Marco}, O. \& {Izzard}, R.~G. 2017, \pasa, 34, e001

\bibitem[{{Decin} {et~al.}(2017){Decin}, {Richards}, {Waters}, {Danilovich},
  {Gobrecht}, {Khouri}, {Homan}, {Bakker}, {Van de Sande}, {Nuth}, \& {De
  Beck}}]{Decin2017}
{Decin}, L., {Richards}, A.~M.~S., {Waters}, L.~B.~F.~M., {et~al.} 2017, \aap,
  608, A55

\bibitem[{{Diamond} {et~al.}(1994){Diamond}, {Kemball}, {Junor}, {Zensus},
  {Benson}, \& {Dhawan}}]{Diamond1994}
{Diamond}, P.~J., {Kemball}, A.~J., {Junor}, W., {et~al.} 1994, \apjl, 430, L61

\bibitem[{{Gail} {et~al.}(2016){Gail}, {Scholz}, \& {Pucci}}]{Gail2016}
{Gail}, H.-P., {Scholz}, M., \& {Pucci}, A. 2016, \aap, 591, A17

\bibitem[{{Gobrecht} {et~al.}(2016){Gobrecht}, {Cherchneff}, {Sarangi},
  {Plane}, \& {Bromley}}]{Gobrecht2016}
{Gobrecht}, D., {Cherchneff}, I., {Sarangi}, A., {Plane}, J.~M.~C., \&
  {Bromley}, S.~T. 2016, \aap, 585, A6

\bibitem[{{Habing} \& {Olofsson}(2003)}]{Habing2003}
{Habing}, H.~J. \& {Olofsson}, H., eds. 2003, {Asymptotic Giant Branch Stars}

\bibitem[{{Herpin} {et~al.}(2006){Herpin}, {Baudry}, {Thum}, {Morris}, \&
  {Wiesemeyer}}]{Herpin2006}
{Herpin}, F., {Baudry}, A., {Thum}, C., {Morris}, D., \& {Wiesemeyer}, H. 2006,
  \aap, 450, 667

\bibitem[{{H{\"o}fner}(2008)}]{Hofner2008}
{H{\"o}fner}, S. 2008, \aap, 491, L1

\bibitem[{{H{\"o}fner} {et~al.}(2016){H{\"o}fner}, {Bladh}, {Aringer}, \&
  {Ahuja}}]{Hoefner2016}
{H{\"o}fner}, S., {Bladh}, S., {Aringer}, B., \& {Ahuja}, R. 2016, \aap, 594,
  A108

\bibitem[{{H{\"o}fner} \& {Freytag}(2019)}]{Hoefner2019}
{H{\"o}fner}, S. \& {Freytag}, B. 2019, \aap, 623, A158

\bibitem[{{H{\"o}fner} \& {Olofsson}(2018)}]{Hoefner2018}
{H{\"o}fner}, S. \& {Olofsson}, H. 2018, \aapr, 26, 1

\bibitem[{{Homan} {et~al.}(2017){Homan}, {Richards}, {Decin}, {Kervella}, {de
  Koter}, {McDonald}, \& {Ohnaka}}]{Homan2017}
{Homan}, W., {Richards}, A., {Decin}, L., {et~al.} 2017, \aap, 601, A5

\bibitem[{{Ireland} {et~al.}(2004){Ireland}, {Tuthill}, {Bedding}, {Robertson},
  \& {Jacob}}]{Ireland2004}
{Ireland}, M.~J., {Tuthill}, P.~G., {Bedding}, T.~R., {Robertson}, J.~G., \&
  {Jacob}, A.~P. 2004, \mnras, 350, 365

\bibitem[{{Ishitsuka} {et~al.}(2001){Ishitsuka}, {Imai}, {Omodaka}, {Ueno},
  {Kameya}, {Sasao}, {Morimoto}, {Miyaji}, {Nakajima}, \&
  {Watanabe}}]{Ishitsuka2001}
{Ishitsuka}, J.~K., {Imai}, H., {Omodaka}, T., {et~al.} 2001, \pasj, 53, 1231

\bibitem[{{J{\"a}ger} {et~al.}(2003){J{\"a}ger}, {Dorschner}, {Mutschke},
  {Posch}, \& {Henning}}]{Jager2003}
{J{\"a}ger}, C., {Dorschner}, J., {Mutschke}, H., {Posch}, T., \& {Henning}, T.
  2003, \aap, 408, 193

\bibitem[{{Jura} \& {Kleinmann}(1992)}]{Jura1992}
{Jura}, M. \& {Kleinmann}, S.~G. 1992, \apjs, 83, 329

\bibitem[{{Kami{\'n}ski} {et~al.}(2017){Kami{\'n}ski}, {M{\"u}ller}, {Schmidt},
  {Cherchneff}, {Wong}, {Br{\"u}nken}, {Menten}, {Winters}, {Gottlieb}, \&
  {Patel}}]{Kaminski2017}
{Kami{\'n}ski}, T., {M{\"u}ller}, H.~S.~P., {Schmidt}, M.~R., {et~al.} 2017,
  \aap, 599, A59

\bibitem[{{Kami{\'n}ski} {et~al.}(2016){Kami{\'n}ski}, {Wong}, {Schmidt},
  {M{\"u}ller}, {Gottlieb}, {Cherchneff}, {Menten}, {Keller}, {Br{\"u}nken},
  {Winters}, \& {Patel}}]{Kaminski2016}
{Kami{\'n}ski}, T., {Wong}, K.~T., {Schmidt}, M.~R., {et~al.} 2016, \aap, 592,
  A42

\bibitem[{{Karovicova} {et~al.}(2013){Karovicova}, {Wittkowski}, {Ohnaka},
  {Boboltz}, {Fossat}, \& {Scholz}}]{Karovicova2013}
{Karovicova}, I., {Wittkowski}, M., {Ohnaka}, K., {et~al.} 2013, \aap, 560, A75

\bibitem[{{Kervella} {et~al.}(2016){Kervella}, {Homan}, {Richards}, {Decin},
  {McDonald}, {Montarg{\`e}s}, \& {Ohnaka}}]{Kervella2016}
{Kervella}, P., {Homan}, W., {Richards}, A.~M.~S., {et~al.} 2016, \aap, 596,
  A92

\bibitem[{{Kervella} {et~al.}(2015){Kervella}, {Montarg{\`e}s}, {Lagadec},
  {Ridgway}, {Haubois}, {Girard}, {Ohnaka}, {Perrin}, \&
  {Gallenne}}]{Kervella2015}
{Kervella}, P., {Montarg{\`e}s}, M., {Lagadec}, E., {et~al.} 2015, \aap, 578,
  A77

\bibitem[{{Khouri} {et~al.}(2014{\natexlab{a}}){Khouri}, {de Koter}, {Decin},
  {Waters}, {Lombaert}, {Royer}, {Swinyard}, {Barlow}, {Alcolea}, {Blommaert},
  {Bujarrabal}, {Cernicharo}, {Groenewegen}, {Justtanont}, {Kerschbaum},
  {Maercker}, {Marston}, {Matsuura}, {Melnick}, {Menten}, {Olofsson},
  {Planesas}, {Polehampton}, {Posch}, {Schmidt}, {Szczerba}, {Vandenbussche},
  \& {Yates}}]{Khouri2014}
{Khouri}, T., {de Koter}, A., {Decin}, L., {et~al.} 2014{\natexlab{a}}, \aap,
  561, A5

\bibitem[{{Khouri} {et~al.}(2014{\natexlab{b}}){Khouri}, {de Koter}, {Decin},
  {Waters}, {Maercker}, {Lombaert}, {Alcolea}, {Blommaert}, {Bujarrabal},
  {Groenewegen}, {Justtanont}, {Kerschbaum}, {Matsuura}, {Menten}, {Olofsson},
  {Planesas}, {Royer}, {Schmidt}, {Szczerba}, {Teyssier}, \&
  {Yates}}]{Khouri2014a}
{Khouri}, T., {de Koter}, A., {Decin}, L., {et~al.} 2014{\natexlab{b}}, \aap,
  570, A67

\bibitem[{{Khouri} {et~al.}(2016){Khouri}, {Maercker}, {Waters}, {Vlemmings},
  {Kervella}, {de Koter}, {Ginski}, {De Beck}, {Decin}, {Min}, {Dominik},
  {O'Gorman}, {Schmid}, {Lombaert}, \& {Lagadec}}]{Khouri2016}
{Khouri}, T., {Maercker}, M., {Waters}, L.~B.~F.~M., {et~al.} 2016, \aap, 591,
  A70

\bibitem[{{Khouri} {et~al.}(2018){Khouri}, {Vlemmings}, {Olofsson}, {Ginski},
  {De Beck}, {Maercker}, \& {Ramstedt}}]{Khouri2018}
{Khouri}, T., {Vlemmings}, W.~H.~T., {Olofsson}, H., {et~al.} 2018, \aap, 620,
  A75

\bibitem[{{Khouri} {et~al.}(2015){Khouri}, {Waters}, {de Koter}, {Decin},
  {Min}, {de Vries}, {Lombaert}, \& {Cox}}]{Khouri2015}
{Khouri}, T., {Waters}, L.~B.~F.~M., {de Koter}, A., {et~al.} 2015, \aap, 577,
  A114

\bibitem[{{Kim} {et~al.}(2018){Kim}, {Cho}, {Yun}, {Choi}, {Yoon}, {Kim},
  {Dodson}, {Rioja}, {Yang}, \& {Yoon}}]{Kim2018}
{Kim}, D.-J., {Cho}, S.-H., {Yun}, Y., {et~al.} 2018, \apjl, 866, L19

\bibitem[{{Kim} {et~al.}(2015){Kim}, {Liu}, {Hirano}, {Zhao-Geisler}, {Trejo},
  {Yen}, {Taam}, {Kemper}, {Kim}, {Byun}, \& {Liu}}]{kim2015}
{Kim}, H., {Liu}, S.-Y., {Hirano}, N., {et~al.} 2015, \apj, 814, 61

\bibitem[{{Kiss} {et~al.}(1999){Kiss}, {Szatm{\'a}ry}, {Cadmus}, \&
  {Mattei}}]{Kiss1999}
{Kiss}, L.~L., {Szatm{\'a}ry}, K., {Cadmus}, Jr., R.~R., \& {Mattei}, J.~A.
  1999, \aap, 346, 542

\bibitem[{{Kozasa} \& {Sogawa}(1997)}]{Kozasa1997}
{Kozasa}, T. \& {Sogawa}, H. 1997, \apss, 251, 165

\bibitem[{{Lawson}(2000)}]{Lawson2000}
{Lawson}, P.~R., ed. 2000, {Principles of Long Baseline Stellar Interferometry}

\bibitem[{{Lebzelter} \& {Hron}(2003)}]{Lebzelter2003}
{Lebzelter}, T. \& {Hron}, J. 2003, \aap, 411, 533

\bibitem[{{Maercker} {et~al.}(2008){Maercker}, {Sch{\"o}ier}, {Olofsson},
  {Bergman}, \& {Ramstedt}}]{Maercker2008}
{Maercker}, M., {Sch{\"o}ier}, F.~L., {Olofsson}, H., {Bergman}, P., \&
  {Ramstedt}, S. 2008, \aap, 479, 779

\bibitem[{{McMullin} {et~al.}(2007){McMullin}, {Waters}, {Schiebel}, {Young},
  \& {Golap}}]{McMullin2007}
{McMullin}, J.~P., {Waters}, B., {Schiebel}, D., {Young}, W., \& {Golap}, K.
  2007, in Astronomical Society of the Pacific Conference Series, Vol. 376,
  Astronomical Data Analysis Software and Systems XVI, ed. R.~A. {Shaw},
  F.~{Hill}, \& D.~J. {Bell}, 127

\bibitem[{{Min} {et~al.}(2009){Min}, {Dullemond}, {Dominik}, {de Koter}, \&
  {Hovenier}}]{Min2009}
{Min}, M., {Dullemond}, C.~P., {Dominik}, C., {de Koter}, A., \& {Hovenier},
  J.~W. 2009, \aap, 497, 155

\bibitem[{{Min} {et~al.}(2003){Min}, {Hovenier}, \& {de Koter}}]{Min2003}
{Min}, M., {Hovenier}, J.~W., \& {de Koter}, A. 2003, \aap, 404, 35

\bibitem[{{Min} {et~al.}(2005){Min}, {Hovenier}, \& {de Koter}}]{Min2005}
{Min}, M., {Hovenier}, J.~W., \& {de Koter}, A. 2005, \aap, 432, 909

\bibitem[{{Mondal} \& {Chandrasekhar}(2005)}]{Mondal2005}
{Mondal}, S. \& {Chandrasekhar}, T. 2005, \aj, 130, 842

\bibitem[{{Nordhaus} {et~al.}(2007){Nordhaus}, {Blackman}, \&
  {Frank}}]{Nordhaus2007}
{Nordhaus}, J., {Blackman}, E.~G., \& {Frank}, A. 2007, \mnras, 376, 599

\bibitem[{{Norris} {et~al.}(2012){Norris}, {Tuthill}, {Ireland}, {Lacour},
  {Zijlstra}, {Lykou}, {Evans}, {Stewart}, \& {Bedding}}]{Norris2012}
{Norris}, B.~R.~M., {Tuthill}, P.~G., {Ireland}, M.~J., {et~al.} 2012, \nat,
  484, 220

\bibitem[{{Ohnaka} {et~al.}(2016){Ohnaka}, {Weigelt}, \&
  {Hofmann}}]{Ohnaka2016}
{Ohnaka}, K., {Weigelt}, G., \& {Hofmann}, K.-H. 2016, \aap, 589, A91

\bibitem[{{Ohnaka} {et~al.}(2017){Ohnaka}, {Weigelt}, \&
  {Hofmann}}]{Ohnaka2017}
{Ohnaka}, K., {Weigelt}, G., \& {Hofmann}, K.-H. 2017, \aap, 597, A20

\bibitem[{{Olofsson} {et~al.}(2002){Olofsson}, {Gonz{\'a}lez Delgado},
  {Kerschbaum}, \& {Sch{\"o}ier}}]{Olofsson2002}
{Olofsson}, H., {Gonz{\'a}lez Delgado}, D., {Kerschbaum}, F., \& {Sch{\"o}ier},
  F.~L. 2002, \aap, 391, 1053

\bibitem[{{Paladini} {et~al.}(2017){Paladini}, {Klotz}, {Sacuto}, {Lagadec},
  {Wittkowski}, {Richichi}, {Hron}, {Jorissen}, {Groenewegen}, {Kerschbaum},
  {Verhoelst}, {Rau}, {Olofsson}, {Zhao-Geisler}, \& {Matter}}]{Paladini2017}
{Paladini}, C., {Klotz}, D., {Sacuto}, S., {et~al.} 2017, \aap, 600, A136

\bibitem[{{Percy} {et~al.}(2001){Percy}, {Wilson}, \& {Henry}}]{Percy2001}
{Percy}, J.~R., {Wilson}, J.~B., \& {Henry}, G.~W. 2001, \pasp, 113, 983

\bibitem[{{Plane}(2013)}]{Plane2013}
{Plane}, J.~M.~C. 2013, Philosophical Transactions of the Royal Society of
  London Series A, 371, 20120335

\bibitem[{{Ridgway} {et~al.}(1982){Ridgway}, {Jacoby}, {Joyce}, {Siegel}, \&
  {Wells}}]{Ridgway1982}
{Ridgway}, S.~T., {Jacoby}, G.~H., {Joyce}, R.~R., {Siegel}, M.~J., \& {Wells},
  D.~C. 1982, \aj, 87, 808

\bibitem[{{Sahai} {et~al.}(2003){Sahai}, {Morris}, {Knapp}, {Young}, \&
  {Barnbaum}}]{Sahai2003}
{Sahai}, R., {Morris}, M., {Knapp}, G.~R., {Young}, K., \& {Barnbaum}, C. 2003,
  \nat, 426, 261

\bibitem[{{Schmid} {et~al.}(2017){Schmid}, {Bazzon}, {Milli}, {Roelfsema},
  {Engler}, {Mouillet}, {Lagadec}, {Sissa}, {Sauvage}, {Ginski}, {Baruffolo},
  {Beuzit}, {Boccaletti}, {Bohn}, {Claudi}, {Costille}, {Desidera}, {Dohlen},
  {Dominik}, {Feldt}, {Fusco}, {Gisler}, {Girard}, {Gratton}, {Henning},
  {Hubin}, {Joos}, {Kasper}, {Langlois}, {Pavlov}, {Pragt}, {Puget}, {Quanz},
  {Salasnich}, {Siebenmorgen}, {Stute}, {Suarez}, {Szul{\'a}gyi}, {Thalmann},
  {Turatto}, {Udry}, {Vigan}, \& {Wildi}}]{Schmid2017}
{Schmid}, H.~M., {Bazzon}, A., {Milli}, J., {et~al.} 2017, \aap, 602, A53

\bibitem[{{Schmid} {et~al.}(2018){Schmid}, {Bazzon}, {Roelfsema}, {Mouillet},
  {Milli}, {Menard}, {Gisler}, {Hunziker}, {Pragt}, {Dominik}, {Boccaletti},
  {Ginski}, {Abe}, {Antoniucci}, {Avenhaus}, {Baruffolo}, {Baudoz}, {Beuzit},
  {Carbillet}, {Chauvin}, {Claudi}, {Costille}, {Daban}, {de Haan}, {Desidera},
  {Dohlen}, {Downing}, {Elswijk}, {Engler}, {Feldt}, {Fusco}, {Girard},
  {Gratton}, {Hanenburg}, {Henning}, {Hubin}, {Joos}, {Kasper}, {Keller},
  {Langlois}, {Lagadec}, {Martinez}, {Mulder}, {Pavlov}, {Podio}, {Puget},
  {Quanz}, {Rigal}, {Salasnich}, {Sauvage}, {Schuil}, {Siebenmorgen}, {Sissa},
  {Snik}, {Suarez}, {Thalmann}, {Turatto}, {Udry}, {van Duin}, {van Holstein},
  {Vigan}, \& {Wildi}}]{Schmid2018}
{Schmid}, H.~M., {Bazzon}, A., {Roelfsema}, R., {et~al.} 2018, ArXiv e-prints

\bibitem[{{Schmidtke} {et~al.}(1986){Schmidtke}, {Africano}, {Jacoby}, {Joyce},
  \& {Ridgway}}]{Schmidtke1986}
{Schmidtke}, P.~C., {Africano}, J.~L., {Jacoby}, G.~H., {Joyce}, R.~R., \&
  {Ridgway}, S.~T. 1986, \aj, 91, 961

\bibitem[{{Szymczak} {et~al.}(1999){Szymczak}, {Cohen}, \&
  {Richards}}]{Szymczak1999}
{Szymczak}, M., {Cohen}, R.~J., \& {Richards}, A.~M.~S. 1999, \mnras, 304, 877

\bibitem[{{Takigawa} {et~al.}(2017){Takigawa}, {Kamizuka}, {Tachibana}, \&
  {Yamamura}}]{Takigawa2017}
{Takigawa}, A., {Kamizuka}, T., {Tachibana}, S., \& {Yamamura}, I. 2017,
  Science Advances, 3, eaao2149

\bibitem[{{van Belle} {et~al.}(1999){van Belle}, {Lane}, {Thompson}, {Boden},
  {Colavita}, {Dumont}, {Mobley}, {Palmer}, {Shao}, {Vasisht}, {Wallace},
  {Creech-Eakman}, {Koresko}, {Kulkarni}, {Pan}, \& {Gubler}}]{vanBelle1999}
{van Belle}, G.~T., {Lane}, B.~F., {Thompson}, R.~R., {et~al.} 1999, \aj, 117,
  521

\bibitem[{{van Leeuwen}(2007)}]{vanLeeuwen2007}
{van Leeuwen}, F. 2007, \aap, 474, 653

\bibitem[{{Vinkovi{\'c}} {et~al.}(2004){Vinkovi{\'c}}, {Bl{\"o}cker},
  {Hofmann}, {Elitzur}, \& {Weigelt}}]{Vinkovic2004}
{Vinkovi{\'c}}, D., {Bl{\"o}cker}, T., {Hofmann}, K.-H., {Elitzur}, M., \&
  {Weigelt}, G. 2004, \mnras, 352, 852

\bibitem[{{Vlemmings} {et~al.}(2017){Vlemmings}, {Khouri}, {O'Gorman}, {De
  Beck}, {Humphreys}, {Lankhaar}, {Maercker}, {Olofsson}, {Ramstedt}, {Tafoya},
  \& {Takigawa}}]{Vlemmings2017}
{Vlemmings}, W., {Khouri}, T., {O'Gorman}, E., {et~al.} 2017, Nature Astronomy,
  1, 848

\bibitem[{{Vlemmings}(2019)}]{Vlemmings2019}
{Vlemmings}, W.~H.~T. 2019, Proceedings IAU Symposium 343: Why galaxies care
  about AGB stars IV

\bibitem[{{Vlemmings} {et~al.}(2003){Vlemmings}, {van Langevelde}, {Diamond},
  {Habing}, \& {Schilizzi}}]{Vlemmings2003}
{Vlemmings}, W.~H.~T., {van Langevelde}, H.~J., {Diamond}, P.~J., {Habing},
  H.~J., \& {Schilizzi}, R.~T. 2003, \aap, 407, 213

\bibitem[{{White} \& {Feierman}(1987)}]{White1987}
{White}, N.~M. \& {Feierman}, B.~H. 1987, \aj, 94, 751

\bibitem[{{Wing}(1971)}]{Wing1971}
{Wing}, R.~F. 1971, \pasp, 83, 301

\bibitem[{{Woitke}(2006)}]{Woitke2006}
{Woitke}, P. 2006, \aap, 460, L9

\bibitem[{{Woodruff} {et~al.}(2009){Woodruff}, {Ireland}, {Tuthill}, {Monnier},
  {Bedding}, {Danchi}, {Scholz}, {Townes}, \& {Wood}}]{Woodruff2009}
{Woodruff}, H.~C., {Ireland}, M.~J., {Tuthill}, P.~G., {et~al.} 2009, \apj,
  691, 1328

\bibitem[{{Young} {et~al.}(1993){Young}, {Phillips}, \& {Knapp}}]{Young1993}
{Young}, K., {Phillips}, T.~G., \& {Knapp}, G.~R. 1993, \apj, 409, 725

\bibitem[{{Zhao-Geisler} {et~al.}(2012){Zhao-Geisler}, {Quirrenbach},
  {K{\"o}hler}, \& {Lopez}}]{Zhao-Geisler2012}
{Zhao-Geisler}, R., {Quirrenbach}, A., {K{\"o}hler}, R., \& {Lopez}, B. 2012,
  \aap, 545, A56

\bibitem[{{Zhao-Geisler} {et~al.}(2011){Zhao-Geisler}, {Quirrenbach},
  {K{\"o}hler}, {Lopez}, \& {Leinert}}]{Zhao-Geisler2011}
{Zhao-Geisler}, R., {Quirrenbach}, A., {K{\"o}hler}, R., {Lopez}, B., \&
  {Leinert}, C. 2011, \aap, 530, A120

\end{thebibliography}

\end{document}